\documentclass[twocol]{ametsoc}

\journal{ercl}

\bibpunct{(}{)}{;}{a}{}{,}

\newcommand{\mtthresh}{T_{\rm thresh}}
\newcommand{\tthresh}{$\mtthresh$}
\newcommand{\tninety}{\ensuremath{T_{\rm 90}}}

\newcommand{\tdur}{\ensuremath{t_{\rm dur}}}       

\newcommand{\nvox}{\ensuremath{V_{\rm MHWS}}}       
\newcommand{\vox}{\ensuremath{{\rm km^2 \; days}}} 
\newcommand{\maxa}{\ensuremath{A_{\rm max}}}       
\newcommand{\aeff}{\ensuremath{\bar A}}       
\newcommand{\adeg}{\ensuremath{[0.25^\circ]^2}}
\newcommand{\aunit}{\ensuremath{{\rm km^2}}}
\newcommand{\vunit}{\ensuremath{{\rm days \, km^2}}}

\newcommand{\exlat}{26^\circ}
\newcommand{\exlon}{217^\circ}
\newcommand{\exdur}{307~days}
\newcommand{\exarea}{6,500,000\,\aunit}
\newcommand{\exvox}{515,000,000\,\vunit}

\newcommand{\nmhwe}{49,034,524}
\newcommand{\nmhws}{649,475}  
\newcommand{\fmhw}{\ensuremath{f_{\rm MHW}}}  

\newcommand{\classa}{minor}
\newcommand{\classb}{moderate}
\newcommand{\classc}{severe}
\newcommand{\ccut}{{\ensuremath10^8}}  
\newcommand{\numc}{226} 

\newcommand{\mlat}{lat}
\newcommand{\lat}{$\mlat$}
\newcommand{\mlon}{lon}
\newcommand{\lon}{$\mlon$}
\newcommand{\qdeg}{\ensuremath{0.25^\circ}}

\title{The Rapid Rise of Severe
Marine Heat Wave Systems}

    \authors{J. Xavier Prochaska
    \correspondingauthor{University of California, Santa Cruz, 
     1156 High St, Santa Cruz, CA 95064.}
 , Claudie Beaulieu and Katerina Giamalaki}

     \affiliation{
     University of California, Santa Cruz, 
     1156 High St, Santa Cruz, CA 95064}

\email{jxp@ucsc.edu}

    \extraaffil{Moody's Analytics, London, UK}

\abstract{
We introduce a new methodology to study marine
heat waves, extreme events in the sea surface temperature
(SST) of the global ocean.
Motivated by previously large
and impactful marine heat waves and by theoretical 
expectation that the dominant heating processes coherently
affect large regions of the ocean, we introduce a 
methodology from computer vision to construct
marine heat wave systems (MWHSs) -- the
collation of SST extrema in dimensions
of area and time.  
We identify \nmhws~MHWSs in the 37~year
period (1983-2019)
of daily SST records and find that the
duration \tdur\ (days), maximum area \maxa\ (km$^2$), and
total ``volume'' \nvox\ (days km$^2$)
for the majority of MHWSs are well-described
by power-law distributions:
  $\tdur^{-3}, \, \maxa^{-2}$ and $\nvox^{-2}$.
These characteristics confirm SST extrema exhibit
strong spatial coherence that define the formation
and evolution of marine heat waves.
Furthermore,
the most \classc\ MHWSs deviate from
these power-laws and are the dominant
manifestation of marine heat waves:  
extrema in ocean heating are driven by the
$\sim 200$ systems with largest area and
duration.  We further demonstrate that the
previously purported rise in the incidence of
marine heat wave events over the past 
decade is only significant in these 
\classc\ systems.
A change point analysis reveals a rapid increase in days under a severe MHW in most regions of the global ocean over the period of 2000-2005. 
Understanding the origin and impacts
of marine heat waves in the current and future
ocean, therefore, should focus
on the production and evolution of the
largest-scale and longest-duration heating phenomena.
}

\begin{document}

\maketitle

%
\section{Introduction}

With the incontrovertible warming of the Earth's ocean, 
both at the surface and at depth \citep[e.g.][]{roemmich2015,Bulgin2020,Johnson2020},
extreme heating events are increasing both 
in frequency and intensity \citep[see][for a review]{oliver2021}.
Termed marine heat waves (MHWs), analogs to the atmospheric extrema that occur
over land \citep[e.g.][]{perkins2015}, these events represent
periods of elevated and sustained ocean warming relative to the recorded
climatology.  
As climate change continues to increase the temperature of our
ocean's surface, MHWs are likely to become more frequent and
impactful on our ecosystem.

Besides serving as putative signatures of global warming,
MHWs have potentially negative impacts on ocean
life, especially when they come in close contact with 
coastal areas \citep[e.g.][]{smith2023}.
Reported consequences on marine life include harmful algal blooms 
\citep{mccabe2016}, 
shifts in species range \citep[e.g.][]{cavole2016,lenanton2017}, 
and even local extinctions
\citep{straub2022}. 
The influence of MHWs
have also been experienced in the economic sector, with instances of marine heat waves affecting aquaculture or important 
fisheries \citep[e.g.][]{mills2013,barbeaux2020}.
The largest marine heat wave ever recorded, dubbed the ``Warm Blob'', occurred in the Northeast Pacific Ocean between 2013 and 2015. 
Regions of the blob exhibited maximum SST 
that reached 6$^\circ$\,C 
above average, sufficient to 
greatly impact the western
seaboard \citep[e.g.][]{bond2015}. 

The primary observable for quantifying MHWs
is the sea surface temperature
(SST), recorded across the global ocean by remote sensing
satellites supplemented by in-situ observations
\citep[e.g.][]{Reynolds2007}.
With such datasets now spanning nearly
40\,years, one may construct a baseline
of SST measurements and search for excursions 
representing extrema.
To date, the majority of MHW literature have defined these
phenomena on small scales, typically the 
$\qdeg \times \qdeg$ discretization of the Level~4 datasets provided for
SST.  The oft-used methodology of
\cite{hobday+2016}, for example, defines a marine heat
wave event (MHWE) as any \qdeg\ cell with SST exceeding
the 90th percentile climatology for at least 5 consecutive
days.  This and subsequent analyses have explored the 
incidence of such MWHEs across time and 
by region and their potential drivers 
\citep[e.g.][]{oliver2018,Frolicher_al_2018, Holbrook2019,jacox2020,SenGupta2020,Laufkotter2020a, vogt2022}
and recent work has constructed
predictive models based on numerical models and machine-learning approaches 
\citep[e.g.][]{jacox2022,giamalaki2022}.

While the physical mechanisms that generate MHWEs
have not yet been firmly established \citep{oliver2021}, 
the leading forcings are radiative heating and wind.
The latter impacts surface cooling, 
ocean currents (i.e.\ advection), 
and the depth of the mixed layer which in turn modulates SST.
These forcings generally act on large spatial scales
($>100$\,km) and often for relatively long durations
($\gg 5$~days).  
Indeed, the community has identified several
long-term and large heating events (e.g.\ the Warm Blob)
that have had significant impacts on the environment and marine life
throughout an entire ocean basin \citep{Zhu_al_2017,cavole2016,Piatt_al_2020,rogers_al_2021}.
Therefore, we consider MHW
phenomena on larger scales than most of the previous works.

In this paper, we develop a new methodology to characterize
SST extrema with the primary goal
to account for spatial and temporal coherence in MHWs.
Specifically, we introduce marine heat wave 
systems (MHWSs) that are the agglomeration of MHWEs 
coincident in area and time.
Our analysis examines the distributions of sizes and
durations and ``volume'' of these collated systems.
We then show that the largest systems, in space and time,
dominate ocean extremes and exhibit the greatest increase of
occurrence in the warming ocean.

\section{Data and Methods}

\subsection{Dataset}

As with several previous treatments of MHWs,
our analysis starts from the National Oceanic and Atmospheric Administration optimal interpolation SST (NOAA OI SST) dataset \citep{Reynolds2007}, 
a global grid with \qdeg\ angular resolution and daily outputs
spanning nearly four decades (1982-2019).
While there exists a wide range of SST datasets available for
exploring extrema on the ocean surface, this product has several positive characteristics
for such analyses:
  (i) daily cadence;
  (ii) coverage of the full ocean;
  (iii) synthesis of satellite and in-situ sensors to 
  compensate for clouds and compromised measurements;
  (iv) relatively high spatial-resolution (\qdeg);
  and 
  (v) nearly 40 years of continuous coverage.
Together these enable the assessment of MHW phenomena
from spatial scales of $\sim 10$~km to entire basins, and 
on timescales of $\sim 1$~week to multiple years.
At the same time, one should be cautious that the assimilation
and interpolation schema of NOAA~OI may introduce correlations 
on time-scales of several days or on spatial scales 
set by, e.g.,  cloud complexes.
These effects, however, are unlikely to significantly impact the 
large-scale and longer-duration phenomena examined in this study.

\subsection{MHWE}
We follow the definition of MHWE from \citet{hobday+2016}. This definition is based on the SST climatology, which refers to the distribution of SST at a given location on a given day of the year (DOY). As we are interested in extrema, we measure percentiles of the SST distribution to establish a threshold \tthresh\ which defines an extreme SST excursion. We adopt the 90th percentile \tninety,
i.e. $\mtthresh = \tninety$. Given \tninety\ derived from the 
climatology, one defines a MHWE for a given cell as any interval 
where the SST exceeds \tninety\ for 5 or more consecutive
days. Furthermore, sets of MHWEs that occur within 2 days
of one another (e.g.\ a gap of 1 day between
a pair of MHWEs) are collated into a single MHWE.
By construction, every position on the ocean will
have SST values that exceed \tninety\ for $\sim 10\%$ 
of the climatological interval and 
therefore MHWEs may be expected
to span the entire ocean. This definition follows \cite{hobday+2016},
differing only in the period used to define the climatology
where we have adopted the full analysis period instead
of the first 20~years (see supplementary section).

\begin{figure*}[h]
 \centerline{\includegraphics[width=40pc]{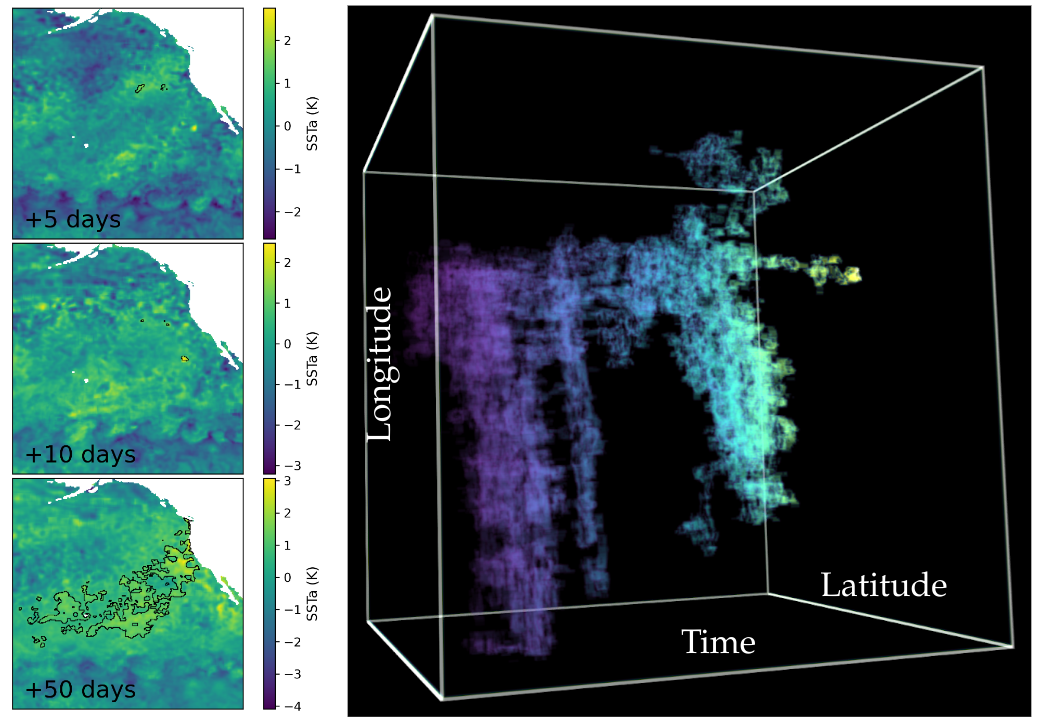}
 }
  \caption{Example of a marine heat wave system (MHWS).
  The left panels show anomalous SST maps 
  for a $70^\circ \times 70^\circ$
  region in the Pacific centered on 
  $\mlat, \mlon \approx \exlat, \exlon$ on the dates of
  1995-10-31,
  1995-11-05,
  and
  1995-12-15
  (top to bottom).
  In each panel, the black contours define the regions that
  satisfy the MHWE definition and contribute to the 
  MHWS, i.e. all have adjacent cells in location or time.
  The right-hand panel is a 3D-rendering of the MHWS where
  the color-coding indicates time (purple to yellow). 
  This MHWS had a duration of $\tdur \approx \exdur$,
  a maximum area 
  $\maxa \approx \exarea$, and a $\nvox \approx \exvox$.
  }
  \label{fig:MHWS_example}
\end{figure*}

\subsection{MHWS}
The basic definition of a MHWS is straightforward:
a MHWS is the collection of all MHWEs that connect in
space and time.
A trivial example is two neighboring MHWEs with 
identical start
and end times, e.g.\ at identical latitude but offset 
by \qdeg\ in longitude.  
These two MHWEs comprise a single MHWS with twice
the area of each MHWE and with identical duration.
Or, if the heating event is dynamic, the MHWS may manifest
first in one location 
of the ocean and `migrate' by connecting
neighboring MHWEs that overlap in time.
Most complex is the connection of
multiple well-separated and long-lived events 
by an intermediate (in time
and space) warming period.

Figure~\ref{fig:MHWS_example} illustrates one MHWS example
from the Pacific Ocean
centered at $\mlat, \mlon \approx \exlat, \exlon$
and with start,end times:
$t_s, t_e = 1995-10-26, 1996-08-27$.
The left panels show anomalous SST (SSTa) 
maps around the region on 
several days between $t_s,t_e$
and highlight cells where the SST values
exceed \tninety.
The right-hand panel shows a rendering of the MHWS
in its three dimensions of space and time.  
We define one ``cube'' in this 3D framework as a voxel; 
it has units of \vox.
This MHWS example, comprised of 44,065~individual MHWEs, 
has a total ``volume'' of 516,746,800~\vox.  
One physical motivation for introducing MHWS's is highlighted
by Figure~\ref{fig:MHWS_example}:  
by December 2015
the MHWS extends many tens of degrees on the ocean 
surface and eventually reaches a maximum area on
a single day of
$\maxa =  6,500,934 \, \aunit$,
indicating a wide-spread forcing mechanism.
Furthermore, the total duration ($\tdur = 307$\,days)
is much longer than the maximum of its constituent 
MHWEs (133~days). 
To understand the origin and impacts of marine
heat waves, it is essential to examine
their total duration and spatial extent.

While statistics on the number and duration of individual
MHWEs in this region might indicate a significant warming
on their own, the construction of a MHWS explicitly captures
the correlations in space and time of the extrema.
Furthermore, the MHWS definition captures the 
translation of extrema across the ocean surface
owing to ocean dynamics or evolving heating processes.  
In practice, the MHWS is the set of connected
voxels in the 3D 
space-time satisfying the MHWE definition.
Every pair of neighboring voxels, in date or location,
is grouped together and the contiguous set
of all voxels defines one MHWS.

The algorithm implemented to construct the MHWSs 
was taken from \cite{shapiro:2001} 
who introduced it 
for computer vision tasks.
We drew inspiration from an application in astronomy
where the authors map Hydrogen emission across 
the sky and in velocity \citep{cantalupo2014}.
We also note the treatment is similar to the methodology of
\cite{Laufkotter2020a} for marine heat waves.
The only significant deviations in our algorithm 
from that of \cite{shapiro:2001}
were  modifications for our spherical geometry and
a restriction that prevents MHWSs 
from spanning across multiple basins.

The primary adjustable parameter for the algorithm
is the minimum number of adjacent voxels required
to define a collection.  In our implementation, 
we have adopted the minimum, i.e. a single 
neighboring pair in space or time.
Experimenting with larger values,  we find 
the primary effect is to eliminate many of the 
minor MHWS.  Increasing this minimum will also 
split apart a number of the MHWS with large \nvox\ 
into multiple, but still large volume MHWS.  
Qualitatively, there is little effect on the 
results or characteristics of the MWHS population 
unless one adopts a value much larger than 1.

When defining MHWSs,  we have chosen to restrict
the spatial collation of 
MHWEs by major ocean basin 
such that a given MHWS is not allowed
to span across more than one basin.
While there may be physical influences that link
extrema in one basin to another, 
our experiments identified
linkages that may be non-physical and we chose
to suppress this behaviour.  
This included one
example that traversed the entire globe with a 
duration of many years.
To enforce the basin separation, we created 
boundaries with a set of lines of constant
latitude or longitude (Table~\ref{tab:basins}).
In practice, this primarily affects MHWSs that 
would have spanned across the eastern Pacific islands
and several that would connect the Indian Ocean
to the Atlantic around Cape Agulhas.


We then calculate a set of 
simple metrics that characterize a MHWS.  
Each voxel has a date $t_i$ and a measured 
area $A_i$ given its location.  
The total duration \tdur\ is the time interval
in days between the earliest time $t_s = min(t_i)$ and 
latest time $t_e = max(t_i)$ for the voxels 
defining the MHWS,

\begin{equation}
  \tdur = t_e - t_s + 1  \;\;\; .
\end{equation}
The maximum area \maxa\ is largest area calculated
for the MHWS on a single day during its duration,
measured in \aunit.

\begin{equation}
    \maxa\ = max(A_{\rm day})
\end{equation}
where 
\begin{equation}
    A_{\rm day}(t) = \sum\limits_i A_i (t_i=t) \;\;\; .
\end{equation}
The MHWS volume \nvox\ is the simple sum of
the volume of each voxel $V_i = A_i t_i$ 
with unit \vox,
\begin{equation}
    \nvox = \sum\limits_i V_i \;\;\; .
\end{equation}
We also consider a characteristic area \aeff\ 
defined as
\begin{equation}
    \aeff \equiv \frac{\nvox}{\tdur} \;\; ,
\end{equation}
which describes the average area occupied by
an MHWS during its duration, measured in \aunit.

Table~\ref{tab:mhws} lists all of the MHWSs
for the fiducial climatology and their salient
properties.  

\subsection{Analysis of MHWS time series}

To characterize and quantify changes in MHWS over 1983-2019, We perform trend detection and 
a change point analysis. A Mann-Kendall (MK) trend test 
\citep{mann1945,kendall1948rank}, 
which is a nonparametric approach, is used to detect trends. For a given time series $X_t, t = 1, 2..., n$, the null hypothesis assumes it is independently distributed, and the alternative hypothesis is that there exists a monotonic trend. The MK statistic is given by

\begin{equation}
\label{eqn:MKstat}
S =  {\sum_{k=1}^{n-1} \sum_{j=k+1}^{n} sign(X_j - X_k)}
\end{equation}

where $X_t$ represents the number of days in a MHW at time $t  (t=1,...,n)$, 
$X_j$ and $X_k$ represent the later-observed and earlier-observed values, respectively $(j > k)$ and

\begin{equation}
    \label{eqn:sign}
    sign(x) = 
    \begin{cases}
    1,  x > 0,\\
    0,  x = 0,\\
    -1, x < 0,
    \end{cases}.
\end{equation}

The statistic $S$ is standardized such that:

\begin{equation}
    \label{eqn:Zstat}
    Z = 
    \begin{cases}
    \frac{(S - 1)} {\sqrt{V(S)}},  S > 0,\\
    0,  S = 0,\\
    \frac{(S + 1)} {\sqrt{V(S)}},  S < 0,
    \end{cases}
\end{equation}

where the variance of $S$ is given by $V(S) = {{[n(n-1)(2n+5) - \sum_{j=1}^{p} t_j(t_j -1)(2t_j +5)]}/18}$ and $t_j$ is the number of data in the tied group and $p$ is the number of groups of tied ranks. The statistic Z follows a standard Normal distribution  with E(Z) = 0 and V(Z) = 1. 

An estimate of the trend is given by the Sen's slope. A set of slopes for all pairs of data that were used to compute S is computed first:

\begin{equation}
\label{eqn:Sen}
d_k =  \frac{(X_j - X_i)}{j-i}
\end{equation}
for $(1 <= i < j <= n)$, where $d_k$ is the slope between data points $X_j$ and $X_i$.
The Sen’s slope $b$ is then calculated as the median from all slopes: $b = median(d_k)$. 

To complement a trend analysis and quantify the timing of the rapid increase in severe MHW days, we perform a change point analysis using the 
non-parametric Pettitt test \citep{pettitt1979} 
due to the non-Gaussian nature of extreme events. Here the null hypothesis is that the time series follow one or more distributions  that have the same location parameter (no
change), against the alternative that a change point exists. The non-parametric statistic is
defined as:

\begin{equation}
\label{eqn:Pettitt}
K_T =  max{|U_{t,T}|},
\end{equation}

where 

\begin{equation}
\label{eqn:Pettitt_U}
U_{t,T} =  \sum_{i=1}^{t} \sum_{j=t+1}^{T} sign(X_i - X_j)
\end{equation}

The most likely timing for a change-point is $K_T$ and its significance is approximated by

\begin{equation}
\label{eqn:Pettitt_pval}
p \approx 2 \exp{ \frac{-6K_T^2}{T^3 + T^2}} 
\end{equation}
Both the Pettitt test and MK trend test are applied using the \texttt{R} package \texttt{trend} 
\citep{pohlert2020}. 




\begin{figure*}[h]
 \centerline{\includegraphics[width=40pc]{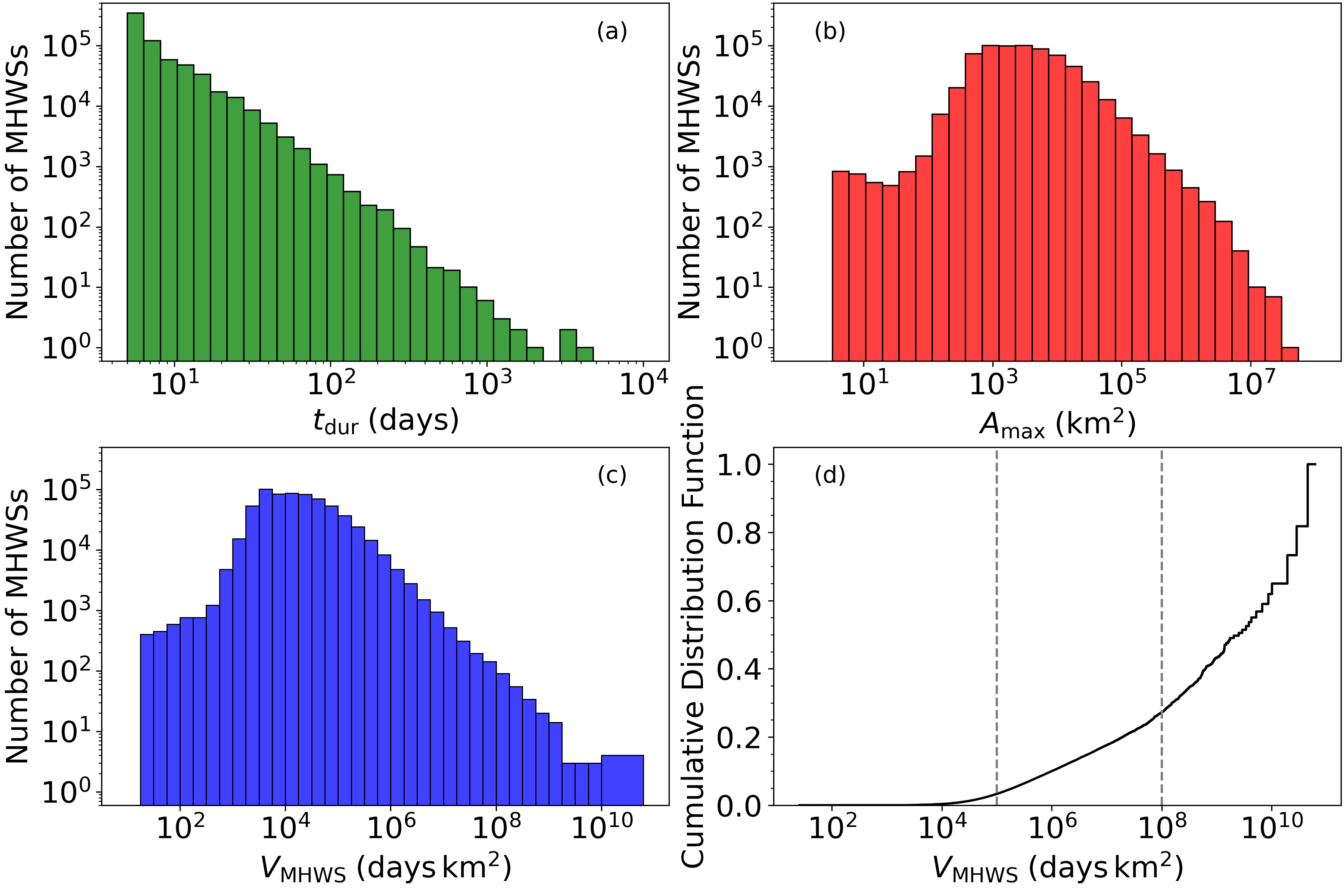}
 }
  \caption{Histograms of the primary characteristics of the MHWS:
  duration \tdur\ (a), 
  maximum area \maxa\ (b),
  and volume \nvox\ (c).
  The \tdur\ distribution is well-described by a power-law
  with exponent $\alpha \approx -3$.
  The \maxa\ and \nvox\ distributions each show a small
  set of systems with low values; these arise primarily 
  in the Arctic and Antarctic Oceans and have negligible
  contribution.  The remainder of these distributions
  are each well-fit by a single power-law with 
  exponents $\alpha \approx -2$.
  At the very highest of \maxa\ and \nvox, however,
  there is a flattening and these \classc\ MHWSs
  dominate the cumulative sum of each distribution.
  This is illustrated in (d)
  which shows the cumulative 
  distribution function of \nvox.
  The dashed lines delineate three classifications of MHWSs:
   (i) \classa,  $\nvox < 10^5$\,vox;
 (ii) \classb,  $10^5 \le \nvox < 10^8$\,vox;
 and
 (iii) \classc,  $\nvox \ge 10^8$\,vox.
  One notes the negligible contribution of the 
  \classa\ MHWS,  the roughly linear contribution 
  of the \classb\ MHWS in logarthmic steps of \nvox,
  and the great contribution of the \classc\ MHWS.
  }
  \label{fig:MHWS_hists}
\end{figure*}

\section{Results}
\label{sec:results}

We have identified and characterized
MHWSs within the 1983-2019 period.
Restricting each MHWS to reside with a single
basin, we recover \nmhws\ unique
MHWSs from \nmhwe\ MHWEs.
The nearly 100$\times$ reduction in the 
number of phenomena demonstrates the strong 
spatial correlation in marine heat waves.
The following subsections present the primary
measurements and results on the MHWSs.

\subsection{{\bf MHWS Properties}}

Figure~\ref{fig:MHWS_hists} presents histograms of the duration \tdur,
maximum area \maxa, and total volume \nvox\ for the full
distribution of \nmhws\ MHWSs from 1983-2019 (inclusive).
The \tdur\ distribution (\ref{fig:MHWS_hists}a)
peaks at the minimum 
value which defines a MHWE (5~days) 
and declines systematically
to the maximum value ($\approx 4,000$ days).
We have fit the \tdur\ distribution with a simple power-law
$\Phi(\tdur) = C \, \tdur^{-\alpha}$ using standard
maximum likelihood techniques for a discrete valued
distribution.  Remarkably, the entire distribution --
spanning three orders-of-magnitude in \tdur\ --
is very well
described by a single power-law with exponent
$\alpha_{t} = -3.0$.  
This `scale-free' distribution is very steep, i.e.\ 
MHWSs are overwhelmingly 
dominated in number by short-term events.
We find approximately $90\%$ of the MHWSs last fewer 
than 15~days.
Despite this, we demonstrate below that these
short-duration MHWSs
contribute negligibly to the integrated extrema on the
ocean's surface.
It is instead
the tail of the \tdur\ distribution that dominates.

Turning to \maxa\ (Figure~\ref{fig:MHWS_hists}b),  one end of the distribution 
shows a small number of MHWS with small
area. These are systems in the polar regions
where \adeg\ corresponds to a small area 
($\approx 100 \, \rm km ^2$);
they are relatively rare and largely inconsequential.
The total distribution peaks at 
$\maxa \approx 10^3 \, \aunit$
and then follows a power-law 
$\Phi(\maxa) \propto \maxa^{-2.1}$ across
four orders of magnitude.  
With an exponent of $\approx -2$, markedly shallower
than the \tdur\ distribution, this implies
that each logarithmic interval of MHWS in \maxa\ covers 
the same integrated area.
Stated another way, the spatial
coherence of SST extrema yields a sufficient 
number of large-area MHWS to
have high impact as extrema on the ocean's surface.

The very largest MHWSs exhibit 
$\maxa > 10^6 \, \aunit$ 
and there are tens of MHWSs
during the period with $\maxa > 10^7 \, \aunit$
(i.e.\ $> 10^3 \, \rm km$ in radius)
or roughly 20\%\ of the Pacific ocean. 
Evidently,
the mechanisms driving marine heat waves
are capable of generating basin-scale extrema.

Lastly, we examine the values of \nvox\ for the MHWSs
which describe the combined duration and sizes
of the systems.
The \nvox\ distribution (Figure~\ref{fig:MHWS_hists}c)
shows similar characteristics as \maxa: 
there is a small set of systems with very
low \nvox\ ($<10^4$\,vox) followed by the 
primary set which tracks a steep power-law
profile.
Analyzing the MHWSs with 
$\nvox > 10^5$\,vox yields the power-law 
$\Phi(\nvox) \propto \nvox^{-1.9}$ which
describes the data well until $\nvox \approx 10^8$\,vox.
As with \maxa, this indicates that the majority
of MHWSs 
occupy the same integrated volume per
logarithmic bin. 
Spatial coherence in SST extrema 
generates very high \nvox\ MHWS that 
contribute significantly to marine heat waves.
Furthermore, 
beyond $\approx 10^8$\,vox there is
yet another flattening in the distribution which
has even greater implications for 
marine heat waves.

The importance of this flattening in $\Phi(\nvox)$
is most
apparent when one considers the total volume
of MHWSs during the entire period.
Figure~\ref{fig:MHWS_hists}d presents the cumulative
contribution of MHWSs to the total as a
function of \nvox.  The systems with lower
values ($\nvox < 10^5$\,vox) dominate in number
but contribute $<5\%$ of the total vox
of the ocean in a marine heat wave state.
We refer to these as `\classa' MHWSs
and interpret them as 
non-impactful fluctuations
just above the 90th percentile in SST.
At $\nvox \approx 10^5$\,vox, the 
distribution transitions
to the $\alpha \approx -2$ power-law
implying the total volume
per logarithmic bin is independent of \nvox,
i.e.\  there is a linear increase in the total vox
with each logarithmic interval in \nvox\
(Figure~\ref{fig:MHWS_hists}d).
We define these as the `\classb' set of MHWSs. 

Last, and most striking, for MHWSs with
$\nvox > 10^8$\,vox,
the total vox per logarithmic bin {\it increases}. 
In fact, these \numc~'\classc' MHWSs comprise 
$\approx 70\%$ of the total ocean designated as satisfying
a marine heat wave condition.
Therefore, the contribution of these \classc\ MHWSs to marine
heat wave extrema exceeds the combined effect of all smaller
systems.
These `severe' MHWSs lie in stark contrast to the main population
and, as demonstrated below, are the dominant phenomena
driving the rise in ocean surface heating.

The single largest MHWS, which includes the infamous
Pacific blob, has persisted in the Pacific Ocean
since the end of 2009.  We provide an animation
of this MHWS (and other \classc\ examples)
which shows it is the agglomeration
of many large structures throughout the full basin.
We expect that these were generated by
several, distinct forcing mechanisms and one might
be inclined to modify the MHWS definition to break
this MHWS into smaller ones.
We emphasize, however, that most of these would 
also satisfy the \classc\ definition and 
the principle conclusions of this work would still hold.

For the remainder of the paper we adopt the following
categories for MHWSs based on the statistics above:
 (i) \classa,  $\nvox < 10^5$\,vox (which corresponds to 85.37\%\ of all MHWS);
 (ii) \classb,  $10^5 \le \nvox < 10^8$\,vox (14.6\%\ of all MHWS);
 and
 (iii) \classc,  $\nvox \ge 10^8$\,vox (0.03\%\ of all MHWS).

 \begin{figure}[h]
 \includegraphics[width=20pc]{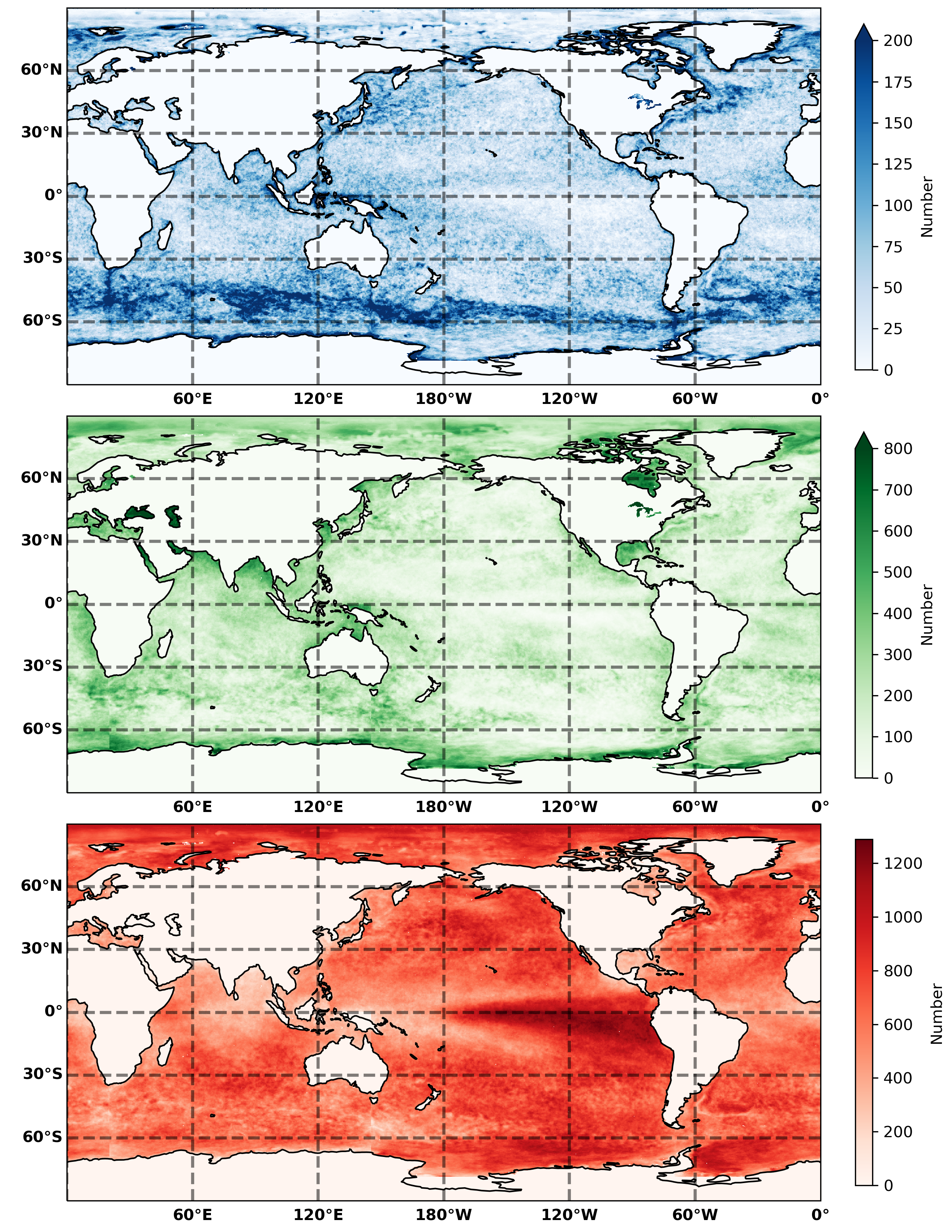}
  \caption{The panels show the number of days that each location
  spent in a MHWS during the full period (1983-2019) in one of the
  three categories: 
  (a) \classa, (b) \classb, (c) \classc.
  Each category shows a progressively higher number of days
  and greater structure in the spatial distribution.
  We identify the highest incidence of \classa\ MHWSs
  within the most dynamic regions of the ocean
  (e.g.\ the Gulf Stream, the ANT).
  In contrast, the \classb\ MHWSs occur primarily
  in coastal regions (e.g.\ Bay of Bengal, Gulf of Mexico).
  Last, the \classc\ MHWSs tend to basin-wide regions 
  and especially the Equatorial Pacific.
  }
  \label{fig:location}
\end{figure}

\subsection{{\bf Geographic Distribution of MHWSs}}
\label{sec:date}

The spatial distribution of MHWS may provide insight
into the forcing mechanisms that generate them and
also highlights portions of the ocean (and coastlines)
most impacted by them.  
Figure~\ref{fig:location}a shows the geographic,
cumulative  distribution of the \classa\ MHWSs 
defined as the number
of days during 1983-2019 that a given location 
exhibited a \classa\ MHWS.
As might be expected, these are roughly uniformly 
distributed across
the ocean with each location exhibiting
an average of $\approx 40$ extrema days over the full period.
The only areas with an excess of these \classa\
MHWSs are the most dynamic regions of the ocean,
e.g.\ the Gulf stream and 
the Antarctic Circumpolar Current (ANT). 
We infer that the complex currents of these regions
lend to larger fluctuations in SST and 
therefore shorter duration MHWSs
\citep[i.e.\ fewer \classb\ and \classc\ systems; see also][]{oliver2021}.
In contrast, the highest latitudes exhibit a 
modest deficit of the \classa\ MHWSs  
(especially the Arctic Ocean).
We believe this is due to the greater spatial coherence in SST
which lends to a higher incidence of large-area \classb\ and \classc\ MHWSs.

A similar map for the \classb\ MHWSs 
(Figure~\ref{fig:location}b)
shows greater geographical variation.  There
is an elevated incidence in these MHWSs near several
coastlines, e.g.\ the Gulf of Mexico, Bay of Bengal, 
Hudson Bay, waters
off western Africa, and the north coast of Australia.
In contrast, the Pacific Ocean and Northern Atlantic
show fewer \classb\ MHWSs. 
Overall, these \classb\ MHWSs track extrema in
modest sized areas with significant
coastlines while generally avoiding the major basins.
Therefore, they may be more impactful on human activity
than regions of the open ocean.


Last, Figure~\ref{fig:location}c reveals the regions
most frequently affected by \classc\ MHWSs.  These
are the Arctic Ocean (which contains
over 20 distinct \classc\ MHWSs), 
the Indian Ocean,
the Pacific Ocean, and the Antarctic Ocean.
The high incidence of \classc\ MHWSs in the Arctic
Ocean reflects recent warming
in that basin but also the relative coherence of SST at 
high latitudes.
The excess of \classc\ MHWSs in the Indian and Pacific
oceans indicate the MHW events in those
regions are especially coherent and extended in duration.
The regions that are avoided by \classc\ MHWSs are
also noteworthy, e.g. the ACC, Bay of Bengal, Gulf of
Mexico.  We suggest the extrema in these areas are 
dominated by local effects and remain relatively 
isolated from the large-scale
circulation and heating of the main basins.

\begin{figure}[h]
 \centerline{\includegraphics[width=20pc]{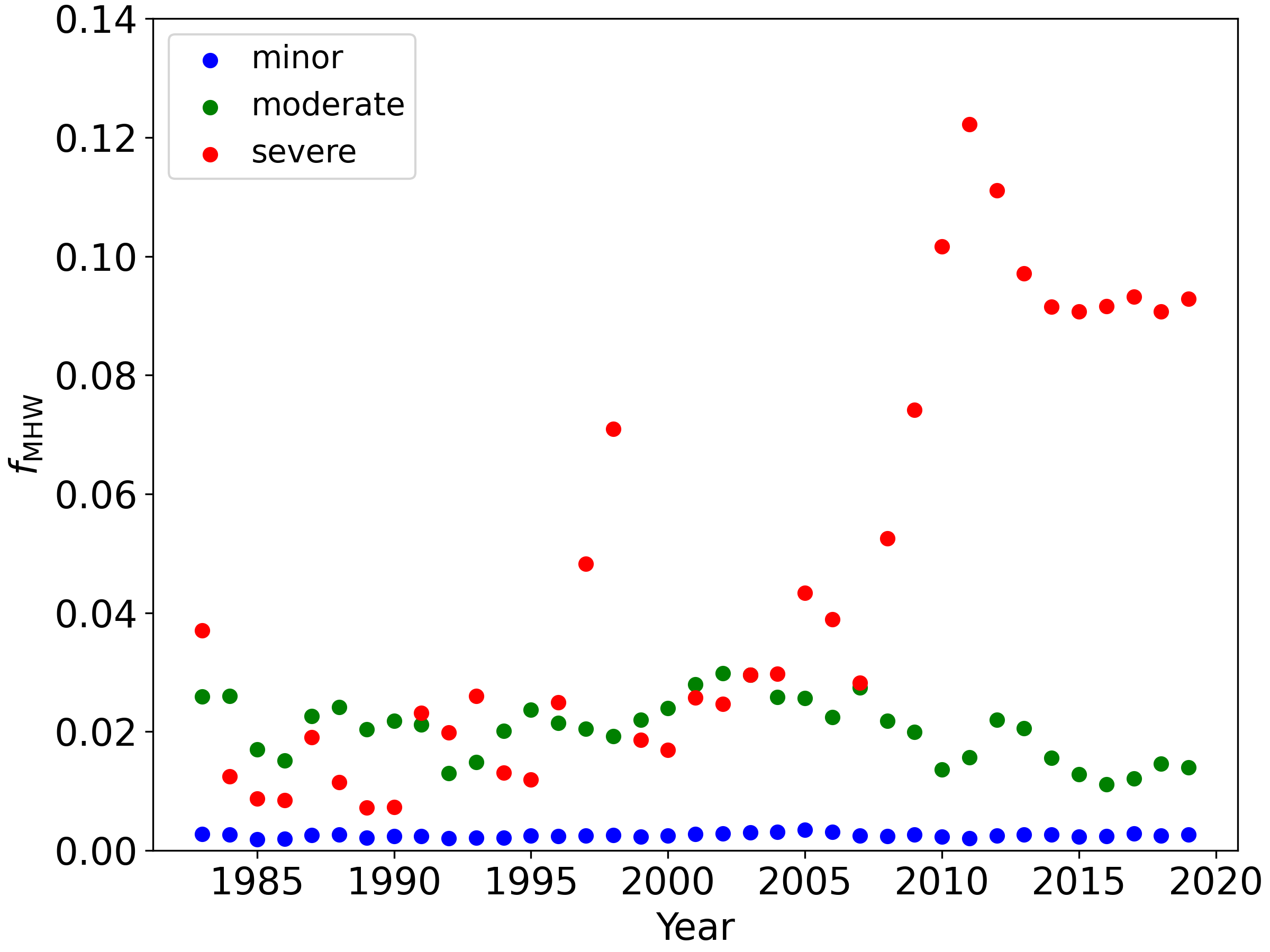}
 }
  \caption{
  Time series for the 
  fraction of the ocean surface 
  \fmhw\ that in a given year exhibits
  one of the three MHWS states: 
  \classa\ (blue), 
  \classb\ (green), 
  \classc\ (red).
  The \classa\ MHWSs affect a very small fraction of 
  the ocean surface ($\ll 1\%$) 
  with little variation in time.
  In contrast, the \classc\ events affect 
  $\approx 10\%$
  and show an $\approx 5\times$ increase over the past
  $\approx 15$\,years.
  }
  \label{fig:global_time}
\end{figure}

\subsection{Time Evolution}
\label{sec:time}

In Figure~\ref{fig:global_time}, 
we explore time evolution in the global incidence of
MHWSs separated by the \classa, \classb, and 
\classc\ categories.
Specifically, we plot the fraction of the ocean surface
in a MHWS each year defined as
$\fmhw = \nvox / (A_{\rm ocean} \times 365 \, \rm days)$
with $A_{\rm ocean}$ the total area of the ocean in
(691,150 cells of \adeg\ or $\approx 364,000,000$~\aunit).
We find the \classa\ MHWS cover less 
than 1\%\ of the ocean per year
and show small variations ($\approx 20\%$) between years. 
Similarly,
the \classb\ MHWS have $\fmhw \approx 2\%$
and also have relatively small temporal 
change across the period, aside from an approximately
$30\%$ decline in the last 10~years.
The most significant trend is the remarkable rise
of \classc\ events after $\approx 2010$, an
$\approx 100-300\%$ increase in \fmhw\ to
cover more than 5\%\ of the ocean surface each year.
Over the past $\approx 10$\,years, the ocean
has exhibited an $\approx 2\times$ increase in 
extrema, with essentially all of this attributed to \classc\ MHWSs
that cover large regions of the ocean for
very long times.

We now quantify changes in MHWS over 1983-2019 and examine 
the presence of rapid increase in \classc\ MHWS
within sub-regions of the ocean. 
We perform trend detection and 
a change point analysis to characterize changes.
The results presented in Figure~\ref{fig:global_time}
imply a significant rise in marine heat waves manifest
in the severe MHWS beginning approximately 15~years ago.

\begin{figure*}[h]
 \centerline{\includegraphics[width=40pc]{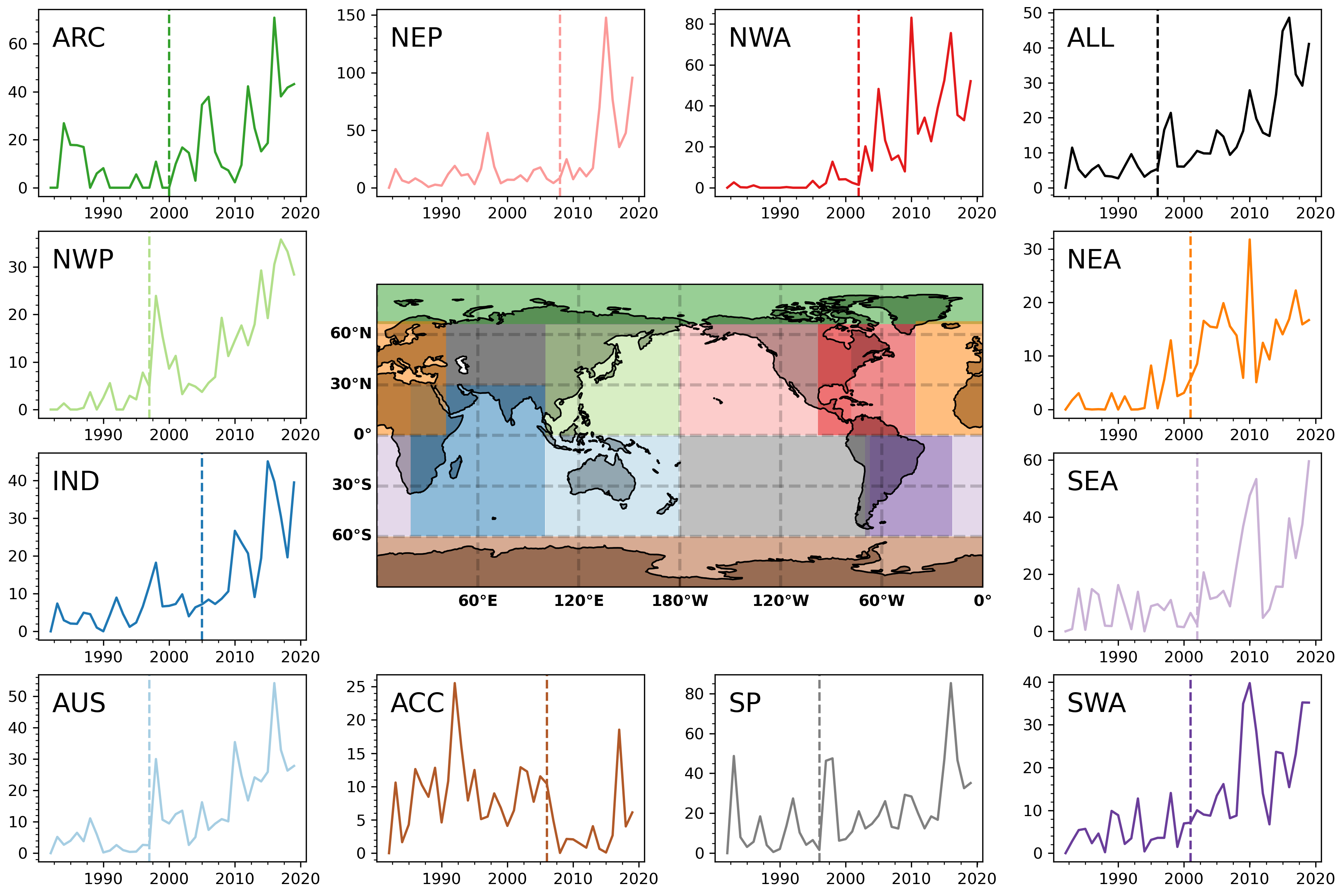}
 }
  \caption{Number of days that each region spent in a severe MHWS during the full period. The map shows the breakdown of regions and their corresponding time series are plotted in the same color. The regions analyzed are 
  (see Table~\ref{tab:regions}): Arctic (ARC), Northwest Pacific (NWP), Indian Ocean (IND), Australian seas (AUS), Northeast Pacific (NEP), Antarctic (ANT), Northwest Atlantic (NWA), South Pacific (SP), Northeast Atlantic (NEA), Southeast Atlantic (SEA), Southwest Atlantic (SWA) and global (ALL). Vertical dashed lines indicate where a changepoint has been detected (Pettitt test, 5\% critical level). 
  In all regions a changepoint is detected and corresponds to a rapid increase in severe MHWS days, except for the Southern Ocean showing a decrease.   }
  \label{fig:change_point}
\end{figure*}

Results are presented in Table~\ref{tab:change}. This analysis reveals a significant ($p < 0.05$) monotonic increase in severe MHW days in 10/11 regions, with the exception of the Antarctic Circumpolar Current (ANT) region that exhibits a significant decrease in severe MHW days over the time period. 

To complement a trend analysis and quantify the timing of the rapid increase in severe MHW days, we apply the Pettitt test to the
total sample (global) and 11~distinct regions 
(Figure~\ref{fig:change_point}).
Specifically, we examined the annual time series
of the number of voxels in a MHWS restricted
to these various geographical regions
(defined by Table~\ref{tab:regions}).

Table~\ref{tab:change} lists the results of this analysis.
We find each region exhibited a statistically
significant ($p < 0.05$) change-point during
the full period. 
The majority of the change points occur
during the years 2000-2005, consistent with our
inferences from Figure~\ref{fig:global_time}. Specifically, an increase in severe MHW days is detected at that time in all regions but the ANT, which is consistent with the results of the Mann-Kendall trend analysis also showing an increase in severe MHW days in the same regions. 

To complement the above investigation,
we also performed a change point analysis on 
the minor and moderate MHWSs, which
Figure~\ref{fig:global_time} suggest have minimal
evolution (or even a modest decline).
Indeed, the change point analysis supports this 
inference with effectively all of the regions
showing very small or negative slopes
(Table~\ref{tab:change}).  Similarly, many of the
change point $p$-values indicate no significant change.
We conclude the temporal evolution
in marine heat wave extrema is driven entirely by 
the \classc\ MHWSs.

\section{Discussion}
\label{sec:discuss}

We have introduced a new definition for marine heat
wave extrema to explicitly allow for and incorporate
spatial coherence within these phenomena.  We have
constructed and characterized \nmhws\ marine heat
wave systems (MWHSs), the collation of individual
marine heat wave events (MHWEs) in space and time,
from \nmhwe\ MHWEs.  
The volume \nvox\ of the MHWSs
ranges from the trivial ($\approx 50 \, \vox$) 
to the extreme (over $10^{10} \, \vox$) with the majority 
following a $\Phi(\nvox) \propto \nvox^{-2}$ power-law
distribution.
We find that the most severe MHWSs -- those with
$\nvox > 10^8 \, \vox$ -- represent $>70\%$ of the ocean
undergoing a marine heat wave extremum.
These severe MHWS have a geographic preference for
the central regions of the major basins.
More importantly, a trend and change point analysis
reveals the severe MHWS have risen rapidly 
to prominence beginning
$\approx 20$~years ago.

A principle implication of the results summarized above
is that marine heat waves exhibit a strong spatial
coherence and that this coherence is common and
impactful.  This may be most evident from the
$\approx 75\times$ agglomeration of MHWEs into MHWSs,
i.e.\ on average 75 MHWEs occupy a common area.
The evidence also includes the shallow
power-law distribution of maximum area \maxa\ 
indicating a significant integrated contribution
from the largest area MHWS.
Finally, 
the majority of the ocean in a
marine heat wave state arises in severe MHWS 
-- the largest, coherent phenomena in the ocean. 
Altogether, we contend the MHWS provide a more natural
description of marine heat wave phenomena than the
more commonly adopted, spatially independent
MHWEs.

\begin{figure}[h]
 \centerline{\includegraphics[width=20pc]{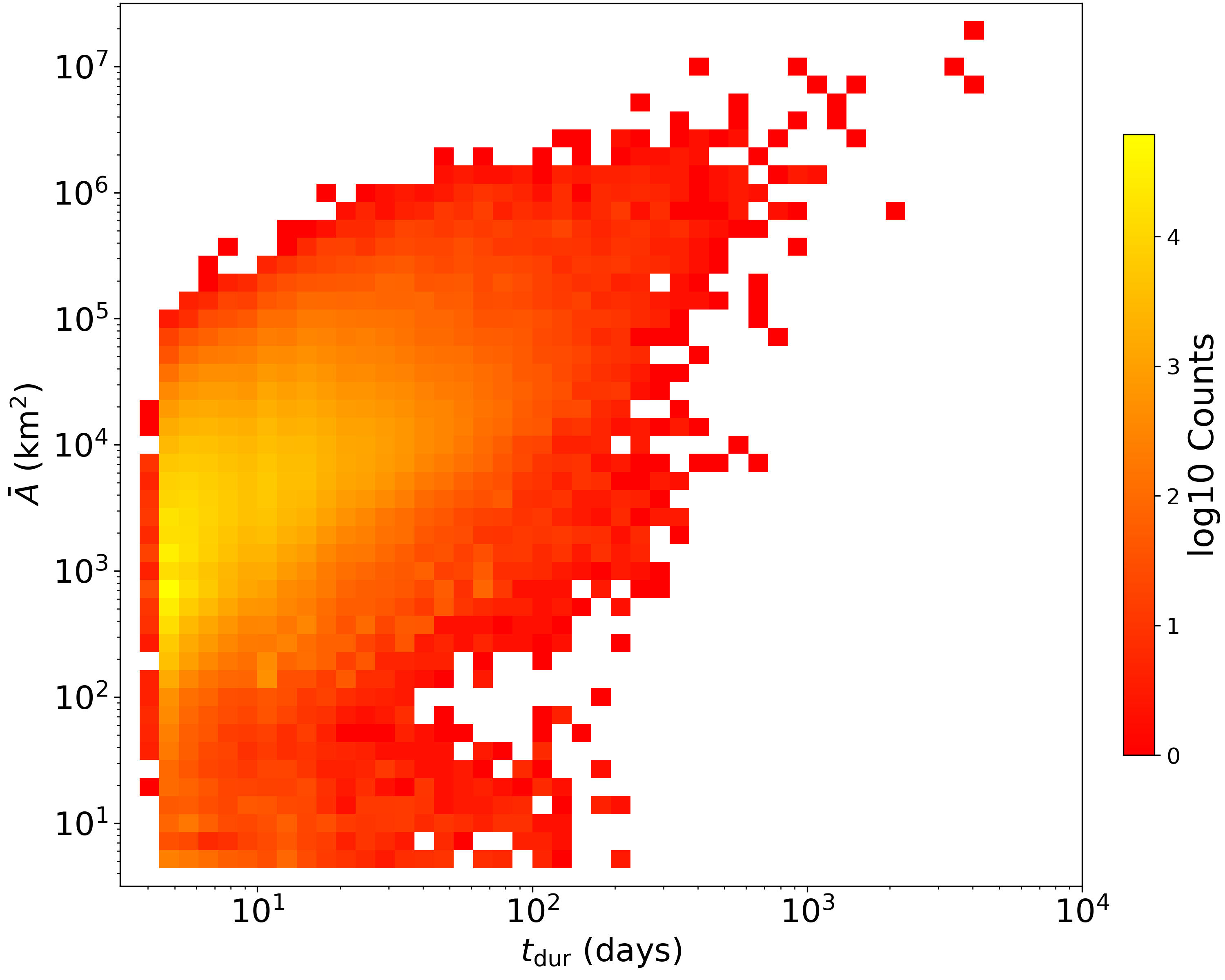}
 }
  \caption{
  2D histogram of the characteristic area
  \aeff\ as a function
  of duration \tdur\ for the full sample of MHWS.
  For systems with durations in excess of 10\,days, 
  nearly all have large areas
  ($> 10^3$\,\aunit), indicating significant spatial
  coherence in marine heat waves.
  The highest areas, which correspond to the 
  severe MHWSs, have $\aeff > 10^7$\,\aunit\ 
  which is an appreciable fraction of an ocean basin.
  These results emphasize the importance of
  spatial extent when characterizing marine heat
  waves and estimating their impacts.
  }
  \label{fig:avg_area_vs_t}
\end{figure}

We further illustrate this point in Figure~\ref{fig:avg_area_vs_t}
which shows the measured characteristic
areas \aeff\ versus the duration \tdur\ of each MHWS.
There is a strong positive correlation between the two
quantities, and
even shorter duration events have significant areas,
e.g. the mean \aeff\ for $\tdur = 50-100$\,days
is 63,500\,\aunit.
At longer durations, one notes that
90\%\ of MHWS with $\tdur > 1$\,year
have $\aeff > 10^5$\,\aunit. 
The extended sizes of marine heat waves are 
fundamental to these extrema, if not their
most defining feature.

\begin{figure}[h]
 \centerline{\includegraphics[width=20pc]{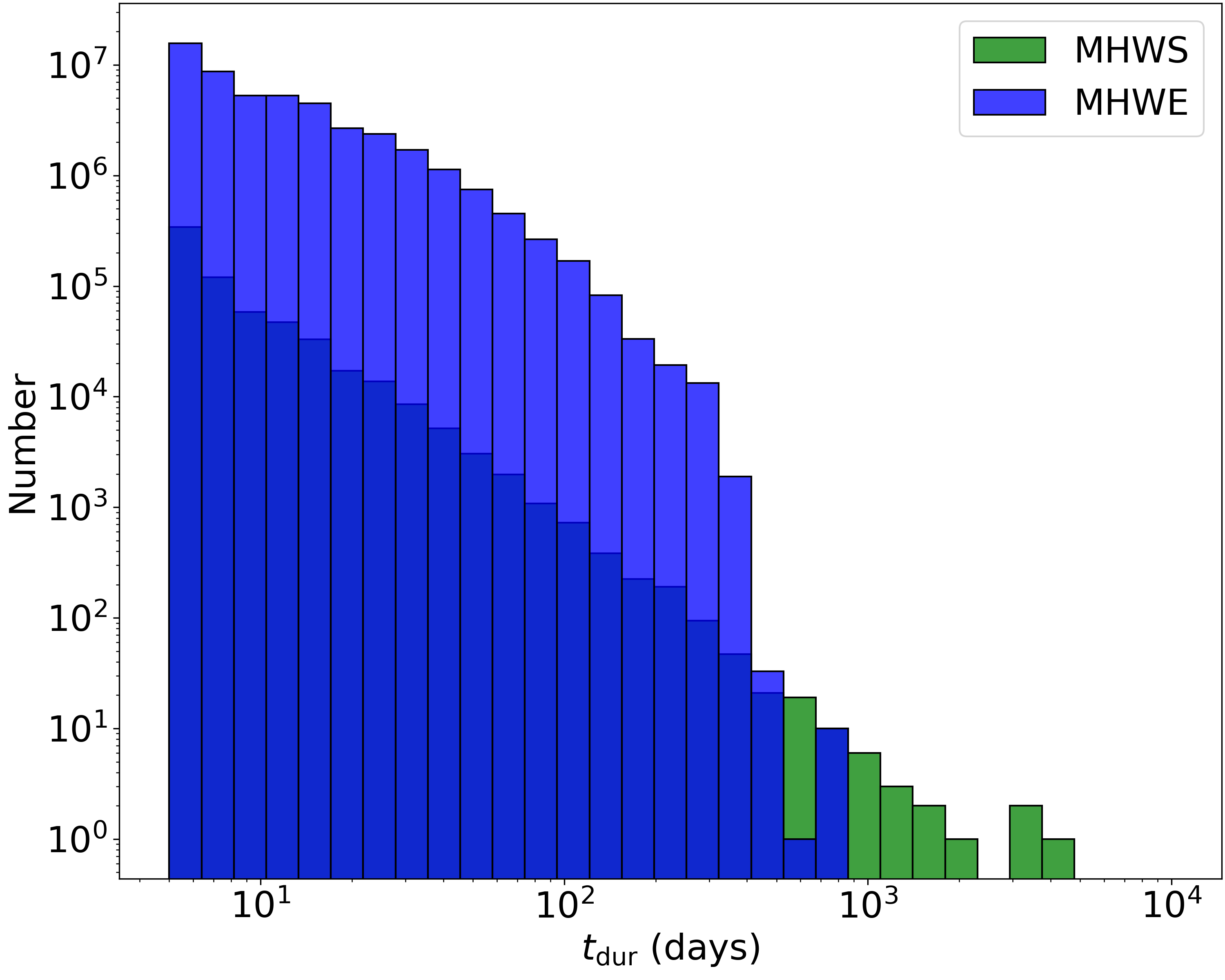}
 }
  \caption{
  Distributions of the durations \tdur\ of 
  marine heat wave events (MHWEs) and 
  marine heat wave systems (MHWSs).
  MHWEs are characterized by short to modest
  durations ($\tdur < 1$\,year).  When one
  accounts for the spatial coherence of 
  these extrema, the majority of these MHWEs 
  are incorporated into 
  larger duration MHWSs with large areas.
  The net effect is that marine heat waves
  are governed by the rare $\tdur >> 1$\,year MHWSs. 
  }
  \label{fig:t_distributions}
\end{figure}

Another implication of these results is that
the spatial coherence of MHWSs
yields durations for the extrema that are qualitatively different
from those of the standard MHWE definition.
This is shown in Figure~\ref{fig:t_distributions}
which compares the \tdur\ distributions of MHWSs
and MHWEs.
Until a duration of 
$\tdur \approx 1$\,year, 
there are $10-100\times$ more MHWE than MHWS
at which point the two distributions cross.
Focusing on MHWEs, one may have concluded that
marine heat waves with durations $\tdur < 1$\,year
have the greatest impacts.
But when one considers the spatial extent of
marine heat waves,  one recovers a 
set with $\tdur >> 1$\,year that 
drive marine heat wave extrema
(e.g.\ Figure~\ref{fig:MHWS_hists}d).
These relatively rare, many-year systems have
the greatest influence on the ocean surface.

\begin{figure}[h]
 \centerline{\includegraphics[width=20pc]{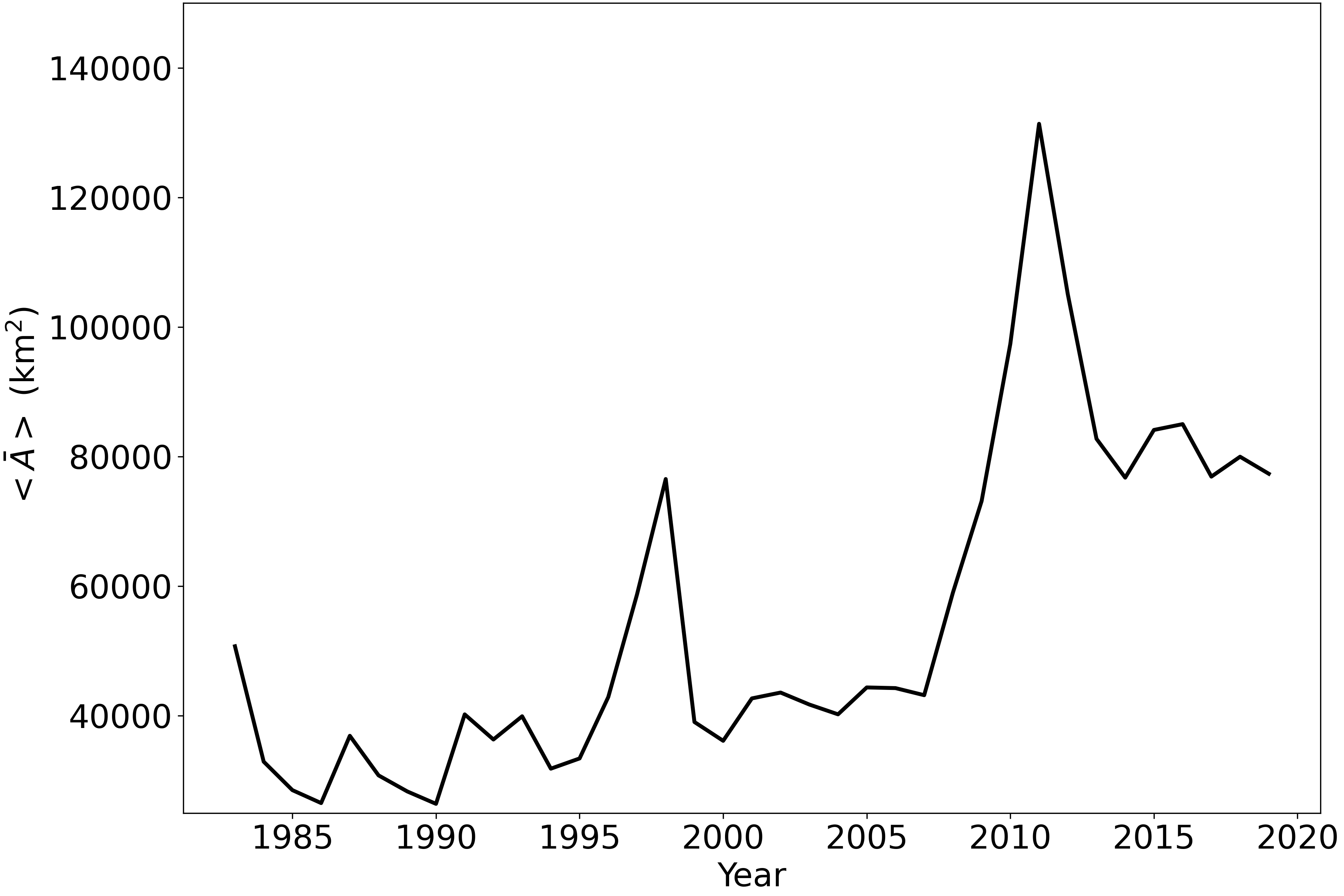}
 }
  \caption{
  Annual averages for the characteristic area $<\aeff>$
  of the MHWSs across the full analysis period.
  Throughout the period, $<\aeff>$ exceeds
  30000\,\aunit, which further emphasizes the strong
  spatial coherence of marine heat waves.
  As important, one observes a sharp rise in 
  $<\aeff>$ near year 2007 to values $3\times$
  higher than the first decade.
  This inflation, which parallels the rise of
  severe MHWSs (Figure~\ref{fig:change_point}),
  signifies a qualitative change in the spatial
  extent of marine heat waves. 
  }
  \label{fig:avg_area_by_year}
\end{figure}

The increase in the fraction of the ocean surface
covered by severe MHWS (Figure~\ref{fig:global_time})
suggests the effective area \aeff\ of the MHWS
is also increasing.
We examine this inference in Figure~\ref{fig:avg_area_by_year},
which plots the annual average of
\aeff\ for all MHWS $<\aeff>$,
with \aeff\ weighted by the number
of days $\Delta t$ the MHWS was active
$<\aeff> = \sum \aeff\ \Delta t / \sum \Delta t$.
Indeed, $<\aeff>$ has increased 
nearly $300\%$ from the first decade to the
most recent.  
The figure lends further evidence
to a recent and sudden rise in severe MHWS,
consistent with the change point analysis of the 
previous section (Figure~\ref{fig:change_point}).

The results presented in 
Figure~\ref{fig:change_point}, and supported by  
Figures~\ref{fig:global_time}
and \ref{fig:avg_area_by_year},
indicate that nearly the entire ocean has
undergone a transition to one where severe 
MHWS (i.e.\ large-area, long-duration) 
are the primary manifestation
of today's marine heat wave extrema.
Our scientific and community focus, therefore,
should be on these few but highly impactful
events.  
Phenomena like the Pacific Warm
Blob of 2013-2015 
\citep{bond2015,tseng2017}
are not one-off, 
oddball events but the new normal. 
We argue the principle focus of future marine heat
wave analysis should be on 
the growth, evolution, and implications of 
these severe MHWS.

Other, previous works have also examined the large
areas of marine heat wave extrema
\citep{Frolicher_al_2018,SenGupta2020}.
Similar to our
MHWS definition, they aggregated neighboring cells
on the ocean surface to measure the contiguous 
areal extent but restricted the aggregation
to individual days.
\cite{Frolicher_al_2018} reported a modest
increase ($\sim 20\%$) in the largest area MHWs when
dividing their 1982-2016 analysis period in half.
This follows the trend displayed 
in Figure~\ref{fig:avg_area_by_year}, although our
results indicate a more rapid rise over the past
15~years and a larger ($3\times$) increase in the average area.
\cite{SenGupta2020}, meanwhile, emphasized a higher
incidence of large area MHWs during El Ni\~no periods
and reported a likely rise in the largest, contiguous
MHW on each given day over the past decade.

More recently, \cite{chapman2022} applied an 
archetype analysis of MHW phenomena in the 
Australasian region to identify large-scale patterns
that underly these extrema.
They then connect these archetypes to teleconnection patterns 
of atmospheric and ocean surface modes.
Future work might explore the archetypes of the 
severe MHWSs presented here to identify climate modes
driving their formation and evolution.
Last, we note the analysis of \cite{sun2023} which
appeared during the review of this manuscript.  Their
approach is similar to the MHWS definition introduced
here in that it connects MHW extrema in both 
space and time, although their analysis focuses 
primarily on tracking the life-cycle and 
movement of individual MHW events.

The extended areas of MHWS have consequences for
the impacts of these extrema on the ocean. 
Consider first the potentially harmful effects
on ocean biology. 
To date, marine heat waves have been implicated for
a range of negative impacts \citep[e.g.][]{smith2023} 
including 
elevated mortality in fish, mammal, and bird populations
\citep{cavole2016,clement2016,oliver2017,straub2022}, 
severe dieback of kelp forests \citep{pearce2013,smale2020}, 
coral bleaching and mortality \citep[e.g.][]{eakin2019},
and the inspiration of harmful 
algal blooms \citep{mccabe2016,trainer2020}.
Of these examples cited above, 
all but one are associated
with a severe MHWS.
Regarding foundation species (e.g.\ marine forests,
coral reefs), 
the large and increasing sizes of MHWSs imply
ever larger regions affected and, likely,
for longer durations.
For some areas and populations, 
one risks complete dieback.
Regarding marine megafauna, 
species that previously could relocate
during a marine heat wave (e.g. whales)
may no longer have that ability.
Last, the larger areas of MHWSs may increase
the size and duration of
harmful algal blooms with severe, knock-on 
consequences for coastal life \cite{trainer2020}.
The ominous implication of Figure~\ref{fig:avg_area_by_year}
is that ever increasing regions of the ocean are
experiencing MHWSs, i.e.\ 
no area may be immune to these extrema.

This also has implications for the tracking and 
prediction of future marine heat waves
\citep[e.g.][]{jacox2022}.
We surmize that the most effective predictor of
where the ocean may next manifest a
marine heat wave is the presence of a nearby, 
severe MHWS.  As an example, 
of the $5.8 \times 10^7$\,\aunit\ of the ocean in
a marine heat wave state on June 1, 2019,
80\%\ of that area was in existing MHWS
on May 1, 2019.
And of these, 98\% were in \classc\ MHWS.
Our results indicate prediction systems should
give preference to existing, long-duration MHWS.
Despite our own modest success with a model that
did not explicitly consider
spatial correlation \citep{giamalaki2022},
we advocate adopting methods that examine the ocean
on much larger scales.

It is worth considering 
the driving factors for the rapid and continued
rise of severe MHWSs.
The over-arching factor is the overall warming of
the global ocean over the past $\sim 20$~years
\citep{Bulgin2020}, inferred to be the result of
anthropogenic influences \cite{oliver2019}.
This has increased the incidence of MHWEs 
and therefore severe MHWSs in nearly all areas 
of the ocean (Figure~\ref{fig:change_point}).
To a large extent, this is the combined effects of
a warming ocean with a fixed baseline climatology, 
even one that spans the full analysis period as
adopted here.
Indeed, adopting a climatology that detrends
the mean reduces the incidence
of severe MHWSs, but these still dominate the
marine heat wave extrema (see supplementary material).
The importance of spatial coherence in marine
heat waves is independent of climatology.


\section{Summary and Concluding Remarks}

Motivated by theoretical and observational evidence
that MHW extrema have large-scale coherence, we
implemented a technique from computer vision to
construct MHWSs, the aggregation of MHW events
in space and time.
We then demonstrated that these MHWS follow power-law
distributions in duration ($\tdur^{-3}$),
maximum area ($\maxa^{-2}$),
and volume ($\nvox^{-2}$).
Based on the distribution of \nvox, we defined three
categories of MHWSs:
\classa,  $\nvox < 10^5$\,vox 
\classb,  $10^5 \le \nvox < 10^8$\,vox 
and
\classc,  $\nvox \ge 10^8$\,vox.
The latter comprise fewer than $0.05\%$ of the MHWS sample
yet deviate from the power-law distribution to 
dominate the MHW extrema.

Furthermore, we find the trends in MHWs 
manifests only in these severe MHWSs and with
a rapid increase in the interval 2000-2005.
This includes all regions of the ocean except
the ANT.
Given these results, we advocate that future MHW analysis
focus primarily on the formation and evolution
of \classc\ MHWs, i.e.\ the
forcing mechanisms that generate and sustain large extrema
for month/year durations.
Of particular interest is whether these are captured
in global circulation models intended to track current
and future climates.

\section{Data Availability Statement}
\label{sec:github}

All of the software developed for this manuscript
is available in GitHub repositories.
For the climatology and MHWE calculations, we 
started with the Python code developed by E. Oliver
(https://github.com/ecjoliver/marineHeatWaves).
We modified these to speed-up performance, testing
during development that we reproduced identical results. 
We then developed alternative climatologies.
Our fork of Dr.\ Oliver's repository is here:
https://github.com/profxj/marineHeatWaves
\citep{Prochaska_Marine_Heat_Waves_2022}.
For MHWSs, we developed a separate repository
found at https://github.com/profxj/mhw\_analysis
\citep{Prochaska_Marine_Heat_Wave_2022}.
This includes all of the Python scripts required
to generate the manuscript figures and tables.

The primary data products of our work are:
  (1) climatologies, stored as individual netcdf files;
  (2) databases of MHWEs, stored as SQL data files and
  {\it parquet} files;
  and
  (3) tables of MHWSs, stored as CSV files.
We also generated simple ``movies'' of several
\classc\ MHWSs as .mov files.
All of these are staged on the dryad long-term
archival system.


\clearpage

\section{Supplementary Materials}
\label{sec:supp}

\subsection{Climatology}
\label{sec:climate}

The climatology in SST analysis generally refers to the mean SST
at a given location on a given day of the year (DOY).
As we are interested in extrema, we measure percentiles
of the SST distribution to establish a threshold \tthresh\ which
defines an extreme SST excursion.
In this analysis, we adopt the 90th percentile \tninety,
i.e. $\mtthresh = \tninety$.

We considered multiple
approaches for calculating \tninety\ before settling on two
for the manuscript: 
  (1) measuring the 90th percentile for the SST distribution at each location over a chosen climatology period;
  and;
  (2) calculating \tninety\ after modifying the SST values
  to remove any linear trend (warming or cooling) in the 
  ocean at that location.  
This ``de-trended'' climatology 
explicitly attempts
to correct for long-term ocean warming, i.e.\ 
it is designed to examine extrema despite 
any such trend.

In both cases, our analysis is
conducted over the full period (1983-2019, inclusive),
where many previous works use only the 
first 20~years of the period (1983-2012 inclusive).
As acknowledged by \cite{hobday+2016}, 
given the steady 
increase in average SST over the past $\sim 15$~years
using a fixed climatological interval
of 1983-2012 lends to a greater incidence of MHWEs in 
recent years.  
For example, in 2019 there are portions 
of the ocean that
nearly exceed \tninety\ throughout the year.

Within the period
one collates the SST measurements from a 11~day interval
centered on the DOY.  This yields 220~values over
the 37~years of the 1983-2019 period modulo 
missing values in the NOAA OI dataset.
The resultant percentile values are then smoothed 
with a running boxcar average of 31~days centered
at the DOY.

\begin{figure*}[h]
 \centerline{\includegraphics[width=40pc]{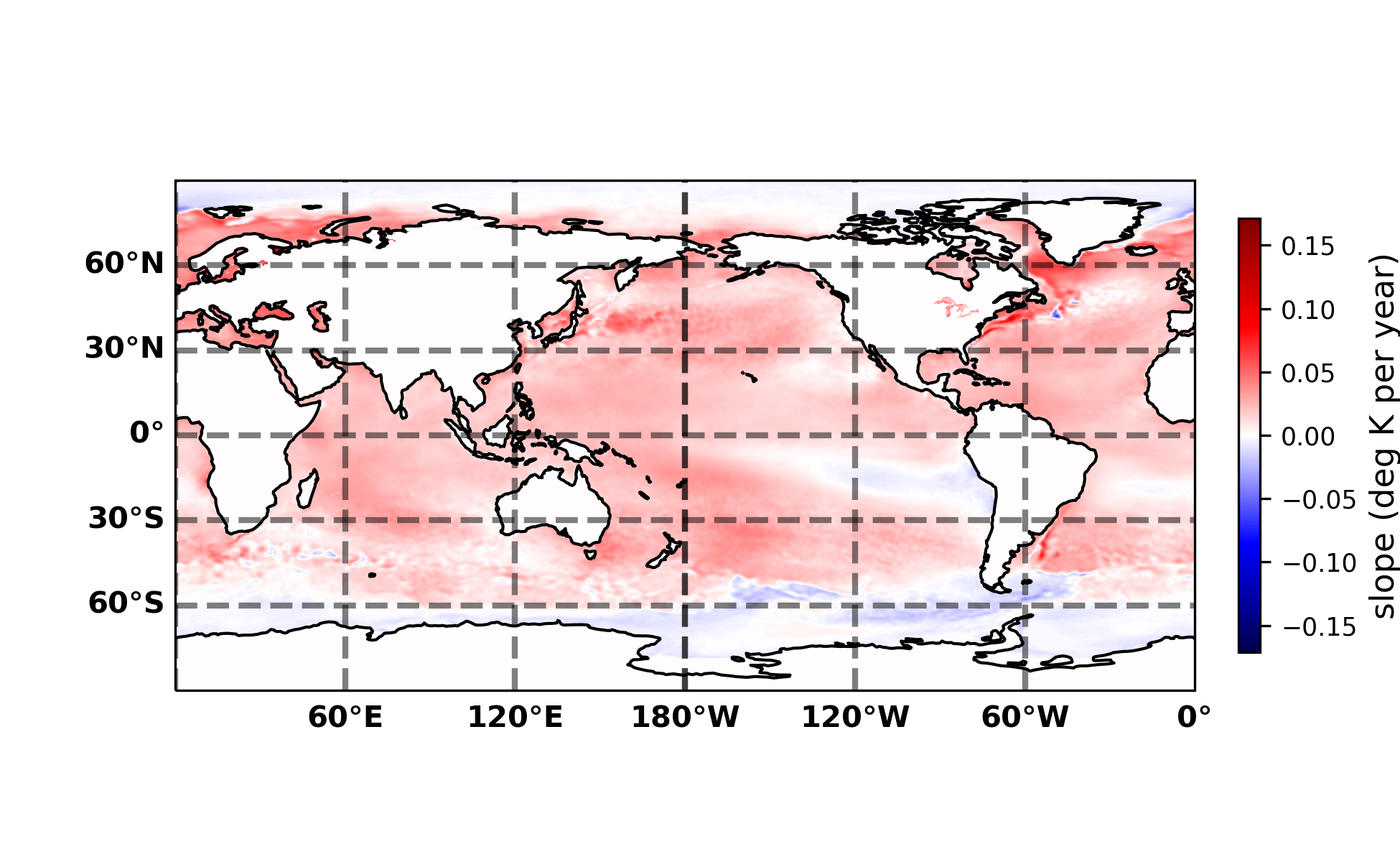}}
  \caption{Measured linear trend in SST evolution from 1983-2019,
  in units of degrees Kelvin per year.  The majority of the ocean shows
  a gradual warming of several hundredths degrees Kelvin
  per year with notable exceptions of the Arctic and Antarctic
  oceans and a portion of the southern Pacific.
  The trend in each cell was removed from the SST values
  to generate the de-trended climatology 
  considered in this manuscript.
  }
  \label{fig:de-trend}
\end{figure*}

Figure~\ref{fig:de-trend} shows the linear trend in SST
($d{\rm SST}/dt$) at each cell measured 
over the full period.  
For the majority of the ocean, 
there is a modest warming of several hundredths
degrees Celsius per year.
Notable exceptions are the Arctic and Antarctic oceans
and waters at lat~$\approx -15^\circ$ in the Pacific Ocean.
For the de-trended climatology, we subtract the 
linear fit from each SST measurement prior to 
assessing \tninety.
We did the de-trending two ways:
  (i) locally, i.e. cell by cell and
  (ii) globally, fitting a trend
  to the median of all cells but only show
  results for the former.

\begin{figure}[h]
 \centerline{\includegraphics[width=20pc]{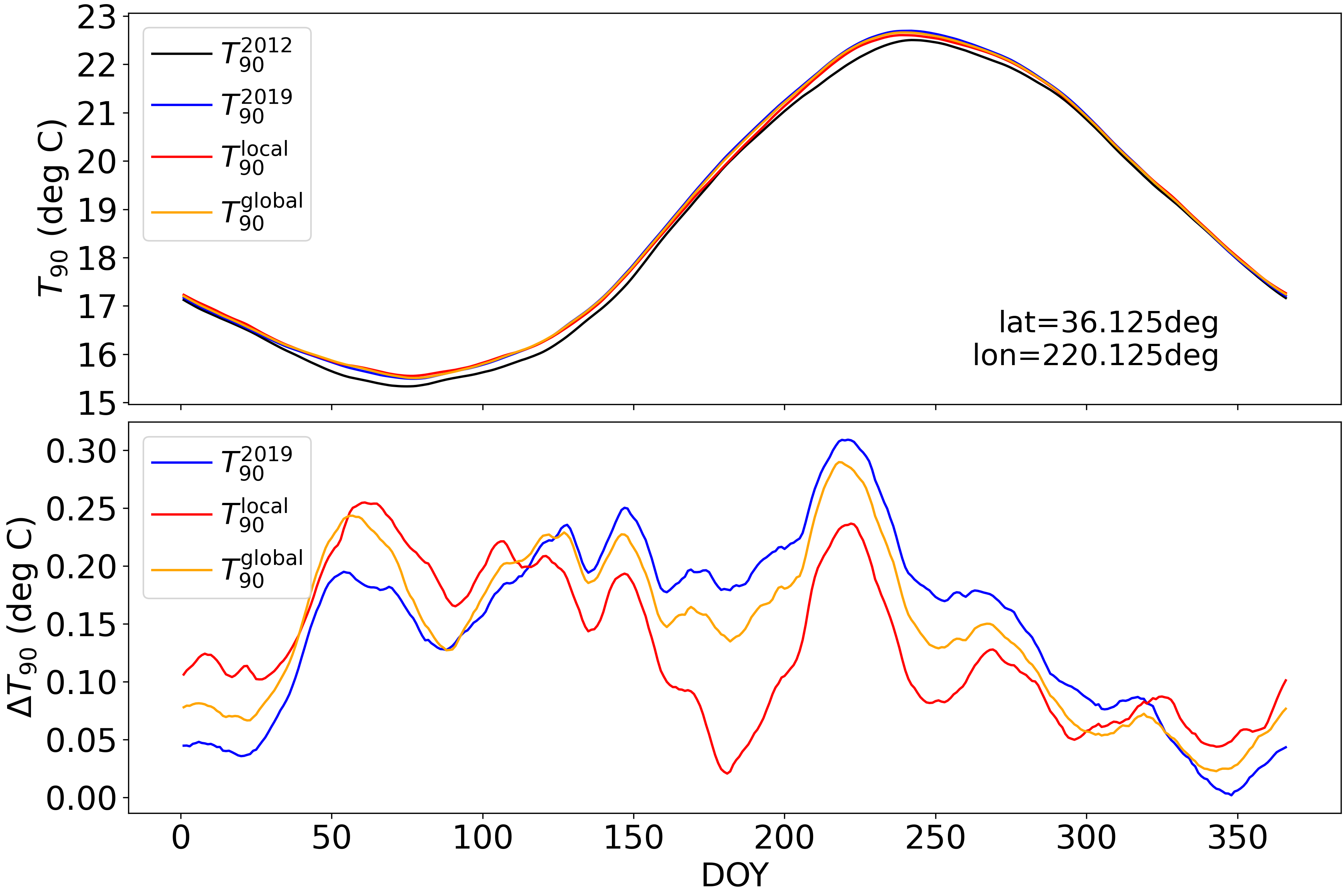}
 }
  \caption{ 
  [{\it upper panel}] Temperature thresholds \tninety\ 
  versus day of year $DOY$
  at an arbitrary location in the northern Pacific.
  The curves show various treatments for 
  defining \tninety: 
    (black) $\tninety^{2012}$, the 90th percentile of the SST distribution on the given $DOY$ based on the 
    1983-2012 climatology;
    (blue) $\tninety^{2019}$, same as above but
    including the years through 2019;
    (red) $\tninety^{\rm local}$, same as 
    $\tninety^{2019}$ but adjusting for
    trend in warming/cooling of the ocean as described
    in Figure~\ref{fig:de-trend} (see text for details).
    (orange) $\tninety^{\rm global}$, also de-trends the SST
    values but from the global median.
    All of these other \tninety\ definitions exceed the 
    measurements from the 1983-2012 climatology. 
  [{\it lower panel}] Difference in the threshold temperature
  $\Delta \tninety$
  relative to $\tninety^{2012}$ for the other three
  climatologies.  Note that the de-trended climatologies
  tend to show smaller $\Delta \tninety$ values because the
  range of corrected SST values has been reduced.
  }
  \label{fig:Tthresh}
\end{figure}

Figure~\ref{fig:Tthresh} compares \tninety\ 
measured from these various approaches to the
climatology
at an arbitrary location in the Northern
Pacific (\lat=36.125~deg, \lon=220.125~deg).  
Extending the climatology period to 2019 yields
higher \tninety\ on each day of the year,
even for the de-trended climatologies.
This leads to fewer extrema at later times in the
period.
One also notes that extending the climatology to 2019
has a significantly greater impact on \tthresh\ than
adjusting for the global warming trend.
However, the SST values are also adjusted in the
de-trended climatology which reduces
MHW phenomena in later years (see below).



 Figure~\ref{fig:NVox_vs_tdur} shows the distribution of
 \nvox\ vs.\ \tdur\ for the full sample of MHWSs and the
 de-trended climatology.  
 As expected, the
 two metrics are highly correlated.  There is, however, a 
 substantial scatter about the main trend, especially for
 the normal set of MHWSs ($\nvox \sim 10^3 - 10^5$\,vox,
 $\tdur \sim 50-300$\,days).
Meanwhile, the \classc\ MHWS all exhibit $\tdur > 100$\,days
with several exceeding 1,000 days.
We have marked on this figure several MHWS discussed
at length in previous literature 
\citep[e.g.][]{Hobday+2018} 
which are naturally recovered by our algorithm.

\begin{figure}[h]
 \centerline{\includegraphics[width=20pc]{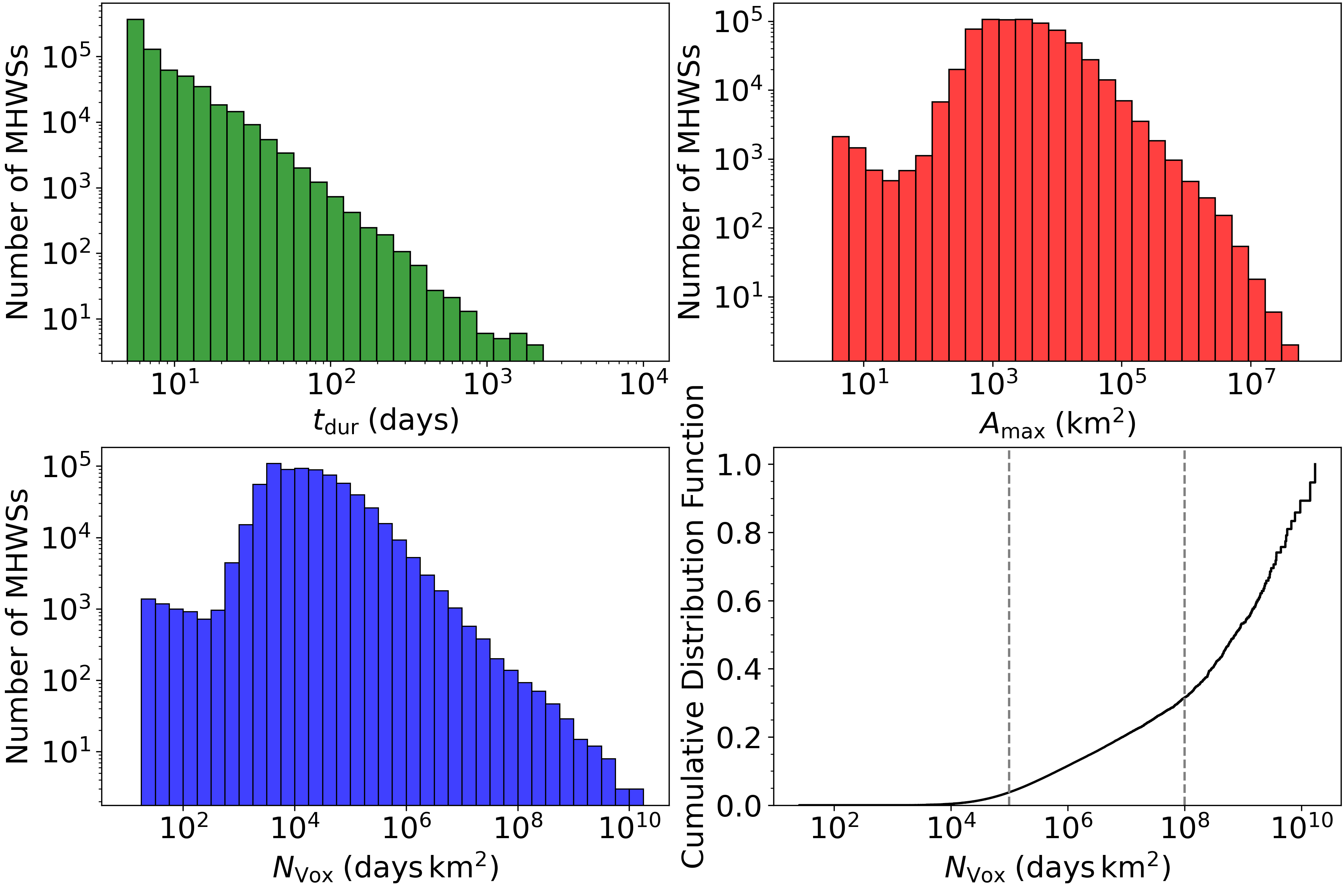}
 }
  \caption{Properties of the MHWS derived with the locally de-trended
  climatology.  Qualitatively, the distributions of 
  duration, maximum area, and volume follow the results for our
  fiducial climatology (Figure~\ref{fig:MHWS_hists}).
  Quantitatively, these are fewer \classc\ MHWS, especially those with
  durations longer than 1~year.
  }
  \label{fig:local_histo}
\end{figure}

\begin{figure}[h]
 \centerline{\includegraphics[width=20pc]{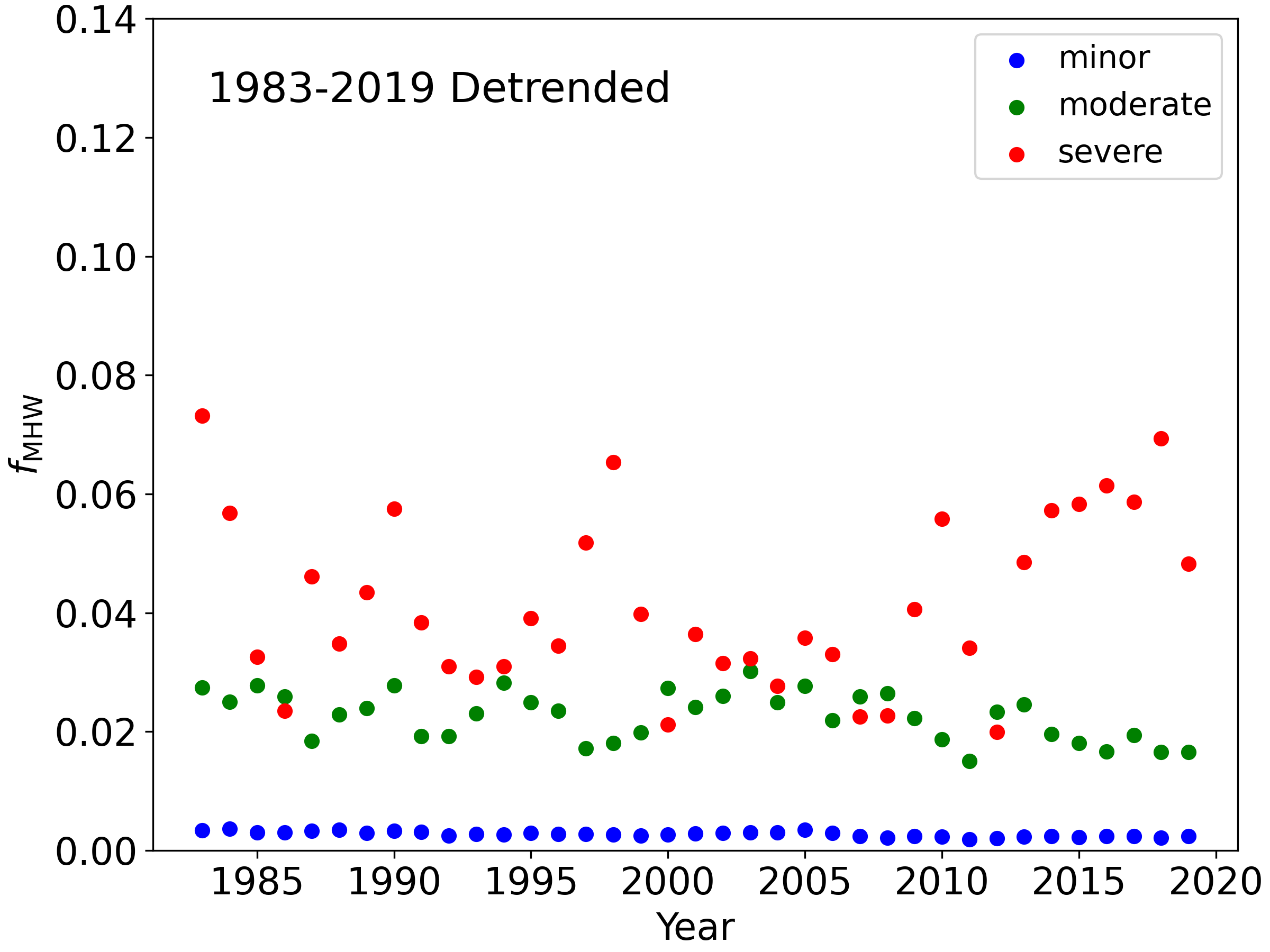}
 }
  \caption{
  Time series for the 
  fraction of the ocean surface 
  \fmhw\ that in a given year exhibits
  one of the three MHWS states: 
  \classa\ (blue), 
  \classb\ (green), 
  \classc\ (red).
  These results are for the locally de-trended
  climatology which yields smaller \fmhw\ and weaker
  trends than the fiducial climatology.
  Nevertheless, one still identifies a rise in \classc\ MHWSs
  over the past decade.  
  This feature indicates the recent increase in marine heat wave
  extrema is non-linear.
  }
  \label{fig:local_Nvox}
\end{figure}

To explore the effects of climatology on the primary
results of this manuscript, we present
Figures~\ref{fig:local_histo} and \ref{fig:local_Nvox}
which show the distributions of MHWS properties and time
evolution for the de-trended climatology.
Qualitatively, the results are similar.
The \tdur, \maxa, and \nvox\ values follow
power-law distributions with exponents very similar
to those of the fiducial climatology.
Furthermore, \classc\ MHWS dominate the total volume
of the ocean that enters a marine heat wave state.
As expected,
the temporal evolution (Figure~\ref{fig:local_Nvox})
does show a smaller rise in incidence over the past
decade but one still identifies the rise in
\classc\ MHWSs over the past decade.


%


\begin{table}[h]
\caption{Definition of Basins Adopted}
\label{tab:basins}
\begin{center}
\begin{tabular}{cccc}
\topline
Basin & Boundary & Location & Range \\
\midline
Indian & E/W & 145E & 90S, 0N \\ 
Indian & E/W & 100E & 0N, 31N \\ 
Indian & E/W & 20E & 90S, 0N \\ 
Indian & S/N & 0N & 145E, 100E \\ 
Pacific & E/W & 70W & 90S, 0N \\ 
Pacific & S/N & 66N & 100E, 120W \\ 
\botline
\end{tabular}
\end{center}
The Boundary separates the basins from East to West (E/W)
or South to North (S/N) at the longitude/latitude Location
given.  The Range of the Boundary is provided by the 
final column.
\end{table}

\begin{table*}[h]
\begin{center}
\caption{Marine Heat Wave Systems}\label{tab:mhws}
\begin{tabular}{ccccccc}
\topline
ID & Lat & Lon & Start & \tdur & \maxa & \nvox \\ 
 & (deg) & (deg) & & (days) & (\aunit) & (days \aunit)  \\ 
\midline
 
1& -78.375& 166.000& 1983-01-05& 5& 311& 1.56e+03\\ 
2& -77.207& 168.606& 1983-03-26& 52& 3.86e+04& 9.42e+05\\ 
4& -78.292& 166.347& 1983-07-22& 8& 1.26e+03& 8.47e+03\\ 
5& -65.114& 189.869& 1983-03-17& 366& 2.74e+06& 1.81e+08\\ 
6& -77.463& 167.973& 1983-09-04& 17& 1.24e+04& 1e+05\\ 
7& -78.327& 166.968& 1983-09-20& 17& 2.19e+03& 2.83e+04\\ 
9& -71.685& 175.861& 1984-03-03& 293& 8.42e+05& 7.03e+07\\ 
11& -78.375& 166.586& 1985-01-06& 6& 1.09e+03& 5.92e+03\\ 
12& -78.138& 166.940& 1985-07-17& 31& 4.5e+03& 8.64e+04\\ 
13& -78.142& 166.616& 1985-08-23& 20& 3.18e+03& 4.97e+04\\ 
14& -77.999& 166.855& 1985-09-15& 46& 5.68e+03& 1.02e+05\\ 
16& -61.980& 194.638& 1985-03-28& 1091& 5.69e+06& 1.39e+09\\ 
19& -78.274& 166.658& 1986-10-06& 35& 1.73e+03& 3.75e+04\\ 
20& -78.375& 165.875& 1987-02-23& 5& 156& 779\\ 
22& -77.614& 166.100& 1987-08-09& 11& 1.14e+04& 8.79e+04\\ 
24& -77.713& 165.316& 1987-12-06& 33& 8.69e+03& 1.77e+05\\ 
25& -75.173& 173.036& 1987-12-05& 94& 1.12e+05& 4.03e+06\\ 
26& -65.714& 248.850& 1988-03-19& 455& 5.5e+06& 6.54e+08\\ 
27& -77.153& 166.967& 1988-09-15& 58& 4.68e+04& 9.36e+05\\ 
29& -78.375& 166.234& 1989-03-06& 6& 623& 3.58e+03\\ 
30& -76.640& 169.068& 1989-03-19& 33& 9.11e+04& 1.6e+06\\ 
31& -75.726& 174.695& 1989-08-23& 203& 3.08e+05& 1.35e+07\\ 
32& -78.375& 166.135& 1990-03-18& 9& 467& 4.05e+03\\ 
33& -71.235& 197.930& 1990-01-22& 323& 1.3e+06& 9.98e+07\\ 
34& -75.777& 182.065& 1990-11-20& 110& 2.94e+05& 9.74e+06\\ 
35& -45.827& 223.795& 1990-12-31& 1211& 2.09e+07& 5.95e+09\\ 
\botline 
\end{tabular}
\end{center}
\end{table*}

\begin{table}[h]
\begin{center}
\caption{Regions for analysis}\label{tab:regions}
\begin{tabular}{ccc}
\topline
Region & Latitudes & Longitudes \\ 
\midline
 
NWP&0N--66N&100E--179W\\ 
AUS&59S--0N&100E--179W\\ 
IND&59S--30N&20E--100E\\ 
ARC&66N--89N&0E--0W\\ 
NEA&0N--68N&39W--41E\\ 
NEP&0N--66N&179W--77W\\ 
NWA&0N--66N&97W--39W\\ 
SEA&59S--0N&17W--20E\\ 
SWA&59S--0N&69W--17W\\ 
SP&59S--0N&179W--66W\\ 
ACC&89S--59S&0E--0W\\ 
\botline 
\end{tabular}
\end{center}
The latitude and longitude ranges define the 11 regions considered in the change point analysis. 
Acronyms are: Arctic (ARC), Northwest Pacific (NWP), Indian Ocean (IND), Australian seas (AUS), Northeast Pacific (NEP), Antarctic (ANT), Northwest Atlantic (NWA), South Pacific (SP), Northeast Atlantic (NEA), Southeast Atlantic (SEA), Southwest Atlantic (SWA). 
For the NEP and NWA regions, we included custom masks to exclude the Gulf of Alaska and Pacific waters, respectively. 
\end{table}

\begin{table}[h]
\begin{center}
\caption{Change point analysis}\label{tab:change}
\begin{tabular}{ccccc}
\topline
Region & Slope & p-value & Changepoint	& p-value \\ 
\midline
 
-- \\
\multicolumn{5}{c}{severe}\\ 
ACC& -0.24& 0.0068& 2006.0& 0.0032\\ 
ALL& 0.67& 4.2e-08& 1996.0& 0.00011\\ 
ARC& 0.64& 0.00085& 2000.0& 0.004\\ 
AUS& 0.74& 2.3e-06& 1997.0& 9.4e-05\\ 
IND& 0.57& 1e-07& 2005.0& 0.00018\\ 
NEA& 0.52& 1.2e-07& 2001.0& 1.1e-05\\ 
NEP& 0.66& 0.00021& 2008.0& 0.0092\\ 
NWA& 1.13& 1.4e-08& 2002.0& 4.4e-06\\ 
NWP& 0.71& 3e-09& 1997.0& 5.4e-05\\ 
SEA& 0.63& 0.00036& 2002.0& 0.0013\\ 
SP& 0.71& 0.00039& 1996.0& 0.015\\ 
SWA& 0.61& 1.6e-06& 2001.0& 0.00011\\ 
\multicolumn{5}{c}{moderate}\\ 
ACC& -0.12& 9.4e-07& 2006.0& 3.5e-05\\ 
ARC& -0.21& 0.00078& 2010.0& 0.026\\ 
AUS& -0.02& 0.61& 2009.0& 0.63\\ 
IND& -0.01& 0.59& 2009.0& 0.72\\ 
NEA& 0.06& 0.0018& 1995.0& 0.015\\ 
NEP& -0.06& 0.0092& 2007.0& 0.066\\ 
NWA& 0.07& 0.12& 1993.0& 0.0061\\ 
NWP& -0.00& 0.84& 2010.0& 0.16\\ 
SEA& -0.12& 0.0086& 2003.0& 0.037\\ 
SP& -0.06& 0.0086& 2005.0& 0.028\\ 
SWA& -0.07& 0.0042& 2006.0& 0.00082\\ 
\multicolumn{5}{c}{minor}\\ 
ACC& -0.01& 6.3e-07& 2006.0& 6.2e-05\\ 
ARC& 0.00& 0.069& 1999.0& 0.026\\ 
AUS& 0.00& 0.95& 1998.0& 0.2\\ 
IND& 0.00& 0.5& 2000.0& 0.34\\ 
NEA& 0.01& 0.0001& 1996.0& 8.2e-05\\ 
NEP& 0.00& 0.74& 2013.0& 0.66\\ 
NWA& 0.01& 0.003& 1997.0& 0.00021\\ 
NWP& 0.00& 0.0035& 1998.0& 0.0012\\ 
SEA& -0.00& 0.2& 1991.0& 0.04\\ 
SP& 0.00& 0.89& 2007.0& 0.89\\ 
SWA& -0.00& 0.12& 2008.0& 0.25\\ 
\botline 
\end{tabular}
\end{center}
\end{table}

\acknowledgments
JXP thanks the University of California for supporting
his research in Oceanography.


 \bibliographystyle{ametsoc2014}
 \bibliography{references}


%



\begin{figure}[h]
 \centerline{\includegraphics[width=20pc]{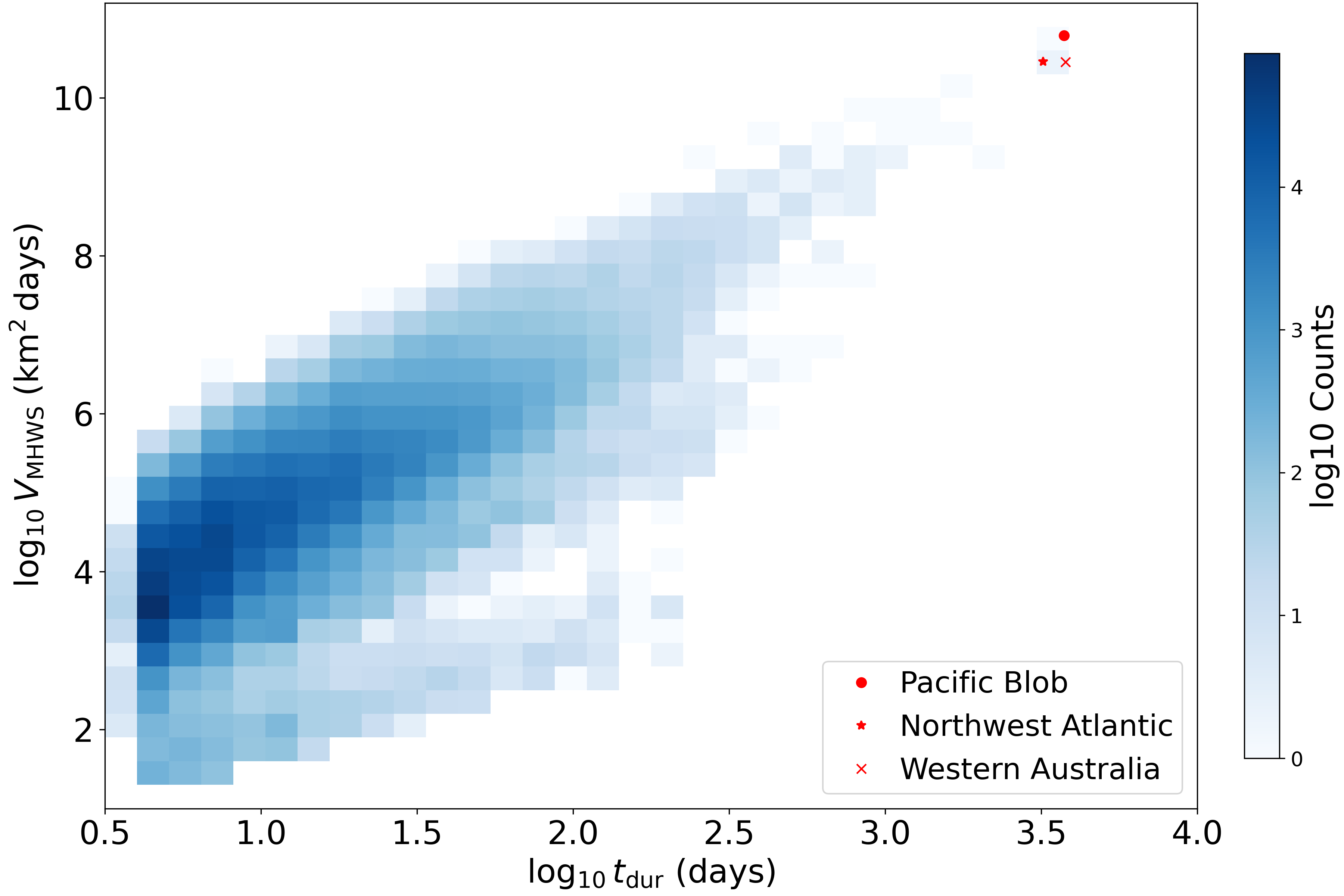}
 }
  \caption{Histogram depicting the relatively tight relation between 
  \nvox\ and \tdur.  The \classc\ MHWS, defined as those with
  $\nvox > \ccut$\,vox, are almost entirely those
  with $\tdur > 300$\,days.  
  Several of the severe MHWSs that contiain prevoiusly studied
  marine heat wave events are 
  marked with red symbols \citep[see][]{Hobday+2018}.
  }
  \label{fig:NVox_vs_tdur}
\end{figure}


\end{document}